\documentclass[12pt,a4paper]{article}
\usepackage{epsf}
\usepackage[dvips]{graphicx}
\usepackage{amsmath, amssymb}
\usepackage{braket}
\usepackage{color}
\usepackage{afterpage}

\begin{document}

\newcommand{\rem}[1]{{\bf #1}}

\renewcommand{\theequation}{\thesection.\arabic{equation}}

\renewcommand{\thefootnote}{\fnsymbol{footnote}}
\setcounter{footnote}{0}

\def \eqref[#1]{Eq.~(\ref{#1})}
\def \figref[#1]{Fig.~\ref{#1}}
\def \secref[#1]{Sec.~\ref{#1}}
\def \partref[#1]{Part~\ref{#1}}
\def \partsecref[#1][#2]{Part~\ref{#1}~Sec.~\ref{#2}}
\def \appref[#1]{Appendix~\ref{#1}}
\def \tableref[#1]{Table~\ref{#1}}
\def \pardif[#1][#2]{\frac{\partial #1}{\partial #2}}
\def \mr[#1]{{\mathrm #1}}
\def \mb[#1]{{\mathbf #1}}
\def \e[#1]{{\mathrm e}^{#1}}
\def \abs[#1]{\left| #1 \right|}

\begin{titlepage}

\def\thefootnote{\fnsymbol{footnote}}

\begin{center}

\hfill UT-13-27\\
\hfill July, 2013\\

\vskip 1.0in

{\Large \bf

Inflationary Gravitational Waves\\
and the Evolution of the Early Universe

}

\vskip .5in

{
Ryusuke Jinno, Takeo Moroi and Kazunori Nakayama
}

\vskip 0.25in

{\em Department of Physics, University of Tokyo,
Tokyo 113-0033, Japan}

\end{center}
\vskip .5in

\begin{abstract}

  We study the effects of various phenomena which may have happened in
  the early universe on the spectrum of inflationary gravitational
  waves. The phenomena include phase transitions, entropy productions
  from non-relativistic matter, the production of dark radiation, and
  decoupling of dark matter/radiation from thermal bath.  These
  events can create several characteristic signatures in the
  inflationary gravitational wave spectrum, which may be direct probes
  of the history of the early universe and the nature of high-energy
  physics.

\end{abstract}

\end{titlepage}

\renewcommand{\thepage}{\arabic{page}}
\setcounter{page}{1}
\renewcommand{\thefootnote}{\#\arabic{footnote}}
\setcounter{footnote}{0}

\tableofcontents

\section{Introduction}
\label{sec_Introduction}
\setcounter{equation}{0}

The early universe is a good laboratory for high-energy physics because the temperature can be much higher than the 
reach of the accelerator experiments.
Thus we may be able to probe high-energy physics by observing some relics of the hot early universe.
The anisotropy of the cosmic microwave background (CMB) carries rich information on the primordial density perturbation,
which is considered to be generated during inflationary era.
Hence the precise observation of CMB gives a clue to inflation models; it is a great success of cosmology for probing high-energy physics.

However, it is rather unknown what have happened in the era between
the inflation and the big-bang nucleosynthesis (BBN).  To go beyond,
one of the promising such relics that carry direct information on the
early universe is the gravitational waves (GWs), since GWs propagate
without interfered by matter and radiation.  In this paper, we focus
on inflationary GWs as a possible probe of the early universe.

The early universe may have experienced several drastic phenomena,
which are not expected in the standard model (SM), e.g., phase
transitions, entropy productions, production of weakly-interacting
particles, etc.  It is known that the spectrum of inflationary GWs
reflects the equation of state of the early universe as studied in
Refs.~\cite{Turner:1990rc,Seto:2003kc,Tashiro:2003qp,Boyle:2005se,Boyle:2007zx,Nakayama:2008ip,Nakayama:2008wy,Kuroyanagi:2008ye,Nakayama:2009ce,Nakayama:2010kt,Schettler:2010dp,Durrer:2011bi,Kuroyanagi:2011fy,Saito:2012bb}.
In particular, it is possible to determine or constrain the reheating
temperature of the
universe~\cite{Nakayama:2009ce,Nakayama:2010kt,Kuroyanagi:2011fy} with
future space laser
interferometers~\cite{Seto:2001qf,Crowder:2005nr,Cutler:2009qv}.  If
there is a brief period of inflation (other than the one responsible
for the present density perturbations) such as thermal
inflation~\cite{Yamamoto:1985rd,Lyth:1995hj}, the spectrum exhibits a
characteristic feature~\cite{Jinno:2011sw}.  The effects of very
weakly interacting relativistic particles, or dark radiation, on the
GW spectrum were pointed out in
Refs.~\cite{Weinberg:2003ur,Dicus:2005rh,Watanabe:2006qe,Ichiki:2006rn,Jinno:2012xb}.
In the presence of decaying matter into dark radiation, the GW
spectrum is deformed in a nontrivial way~\cite{Jinno:2012xb}.

In a realistic setup, the GW spectrum would be affected in a
complicated way.  For example, in the Peccei-Quinn (PQ) model for
solving the strong CP problem~\cite{Peccei:1977hh,Kim:1986ax}, the PQ
phase transition may have occurred at around the cosmic temperature $T
\sim 10^9$--$10^{12}$\,GeV, and huge amount of entropy could have been
released in association with the phase transition.  Relativistic
axions may have been also efficiently produced by the decay of the PQ
scalar condensate, which would contribute as dark radiation.  All
these phenomena significantly affect the spectrum of inflationary GWs.
In other words, it may be possible to access the high-energy phenomena
and underlying physical processes by studying the detailed spectrum of
inflationary GWs.

In this paper we study the inflationary GW spectrum in detail in
various setups having concrete models in mind.  The combinations of
above mentioned effects make several characteristic features in the
spectra, which enable us to infer information on the early universe
phenomena.

First we review basic properties of GWs in Sec.~\ref{sec_Properties},
including the GW spectrum and effects of the anisotropic stress.  In
Sec.~\ref{sec_Illustration}, we exhibit some simple examples of the GW
spectrum as a first step to the following section.  In
Sec.~\ref{sec_Examples_of_possible_GW_spectrum}, we combine all these
effects to show the realistic GW spectra with several typical
examples.  In Sec.~\ref{sec_Model} particle physics motivated models
will be provided in which the GW spectrum actually becomes
complicated.  Sec.~\ref{sec_conc} is devoted to conclusions.

\section{Basic properties of GWs}
\label{sec_Properties}
\setcounter{equation}{0}

\subsection{Background Evolution}

Before discussing the evolution of GWs we first explain that of
background, which we assume throughout this paper to be the
Friedmann-Robertson-Walker (FRW) universe with negligible curvature.
The FRW metric and its tensor perturbation are given by
\begin{eqnarray}
ds^2 = -dt^2 + a^2(t)(\delta_{ij}+h_{ij}(t,\mb[x]))dx^i dx^j,
\label{eq_metric}
\end{eqnarray}
where $a(t)$ is the scale factor and $h_{ij} = h_{ji}$ satisfies transverse and traceless condition: $h_{ii}=h_{ij,i} =0$.

The evolution of the FRW universe is described by the Friedmann equation
\begin{eqnarray}
H^2 = \frac{8 \pi G}{3} \rho_{\rm tot} = \frac{1}{3M_P^2} \rho_{\rm tot},
\label{eq_Friedmann}
\end{eqnarray}
where $H=\dot{a}/a$, and $M_P \simeq 2.4 \times 10^{18}$GeV is the
reduced Planck mass.  The total energy density of the universe,
$\rho_{\rm tot}$, generally include
\begin{equation}
	\rho_{\rm tot} = \rho_{\rm vac} +  \rho_{\rm m} + \rho_{\rm r} + \rho_{X},
\end{equation}
where $ \rho_{\rm vac}, \rho_{\rm m}, \rho_{\rm r}$ and $ \rho_X$
represent the vacuum, matter, (visible) radiation and dark radiation
energy densities, respectively.\footnote
{In this paper, we call the energy density with $w=-1$ (with $w$ being 
the equation-of-state parameter) ``vacuum energy.''}
The radiation energy density $\rho_{\rm r}$ is related to the cosmic
temperature $T$ as
\begin{eqnarray}
\rho_{\rm r}
= \frac{\pi^2}{30} g_{*} T^4,
\end{eqnarray}
where $g_{*}$ is the relativistic degrees of freedom.  Unless
otherwise stated, we use $g_{*} = 228.75$, which is the value in the
minimal supersymmetric (SUSY) standard model (MSSM) at high enough
temperature.\footnote{ Including other fields than in MSSM causes
  deviation from this value, but we neglect it as small.  } The
amounts of other components depend on the particle physics model and
there can be energy transfers among these components.  Thus the Hubble
expansion rate reflects their behavior in the early universe and it
directly affects the GW spectrum as shown below. We will study
concrete setups in the following sections.

\subsection{Evolution of GWs: case without dark radiation}

\subsubsection{Evolution equation}

Substituting the metric \eqref[eq_metric] into the Einstein equation,
we get the equation of motion of GWs:
\begin{eqnarray}
\ddot{h}_{ij}+3H\dot{h}_{ij}-\frac{\nabla^2}{a^2}h_{ij} = 16 \pi G \Pi_{ij},
\label{eq_GW_eom_x}
\end{eqnarray}
where $\Pi_{ij}$ is the anisotropic stress of the energy-momentum tensor, 
which satisfies $\Pi_{ii}=\Pi_{ij,i} =0$.
We decompose $h_{ij}$ using polarization tensors $\mr[e]_{ij}^{+,\times}$,
\begin{eqnarray}
h_{ij}(t,\mb[x]) 
= \sum_{\lambda = +,\times} \int \frac{d^3k}{(2\pi)^3} h(t,\mb[k],\lambda) e^{i\mb[k] \cdot \mb[x]} \mr[e]_{ij}^{\lambda},
\end{eqnarray}
to rewrite \eqref[eq_GW_eom_x] into
\begin{eqnarray}
\ddot{h}(t,\mb[k],\lambda)+3H\dot{h}(t,\mb[k],\lambda)+\frac{k^2}{a^2}h(t,\mb[k],\lambda) 
= 16 \pi G \Pi(t,\mb[k],\lambda).
\label{eq_GW_eom_k}
\end{eqnarray}
Here $k=|\mb[k]|$, and we have performed the same decomposition and
transformation on $\Pi_{ij}$ as is done on $h_{ij}$.  Also, we define
the polarization tensors so that they satisfy $\mr[e]^{\lambda}_{ij} =
\mr[e]^{\lambda}_{ji}, \mr[e]^{\lambda}_{ii} = \mr[e]^{\lambda}_{ij,i}
= 0$ and $\mr[e]^{\lambda}_{ij} \mr[e]^{\lambda'*}_{ij} =
\delta_{\lambda \lambda'}$.

For later use, we introduce the variable $u$ defined by
\begin{eqnarray}
  u = k\eta = k\int_0^t \frac{dt'}{a(t')},
\end{eqnarray}
and rewrite \eqref[eq_GW_eom_k] to get
\begin{eqnarray}
h''(u,\mb[k],\lambda)+2H_u h'(u,\mb[k],\lambda)+h(u,\mb[k],\lambda) 
= 16 \pi G \left( \frac{a}{k} \right)^2 \Pi(u,\mb[k],\lambda),
\label{eq_GW_eom_u}
\end{eqnarray}
where the prime denotes the derivative with respect to $u$ and $H_u
\equiv a'/a$.  If $a(t) \propto t^p$ as in the radiation dominated
(RD) era $(p=1/2)$ or the matter-dominated (MD) era $(p=2/3)$, we
obtain $u = \frac{p}{1-p} \frac{k}{aH}$.  Imposing the initial
condition of $h(t,\mb[k],\lambda) \to h_{\rm prim}(\mb[k],\lambda)$
for $t\to 0$, the solution of \eqref[eq_GW_eom_u], when the
anisotropic stress is neglected, is given by
\begin{equation}
	h(u,\mb[k],\lambda) =h_{\rm prim}(\mb[k],\lambda)  j_0(u) ~~~{\rm for~RD},
\end{equation}
and
\begin{equation}
	h(u,\mb[k],\lambda) = h_{\rm prim}(\mb[k],\lambda) \frac{3j_1(u)}{u} ~~~{\rm for~MD},
	\label{h_MD}
\end{equation}
where $j_i$ are the $i$-th spherical Bessel function:
\begin{equation}
	j_0(u) = \frac{\sin(u)}{u},~~~\frac{j_1(u)}{u} = \frac{\sin(u) - u\cos(u)}{u^3}.
\end{equation}
From this solution, it is easily seen that $h(u,\mb[k],\lambda) \sim {\rm const.}$ for the modes outside the horizon 
$(k \ll aH)$ and $h(u,\mb[k],\lambda) \propto a^{-1}$ for the modes inside the horizon $(k \gg aH)$.

\subsubsection{GW spectrum: modes entering the horizon at the RD era}

Here we define basic quantities used in the following sections.  The
energy density of the GWs is given by (see Appendix \ref{app:tensor}
for details)
\begin{eqnarray}
  \rho_{\rm GW}(t)
  = \int d\ln k \rho_{\rm GW} (t,k)
  = \frac{1}{32\pi G} \braket{h^{ij;0} h_{ij;0}}_{\rm osc},
\end{eqnarray}
where $\rho_{\rm GW}(t,k)$ denotes the energy density of tensor
perturbation per logarithmic frequency. In addition,
$\braket{\cdots}_{\rm osc}$ denotes the oscillation average (and hence
the above expression is relevant only for the sub-horizon mode).
Taking the ensemble average, we obtain
\begin{eqnarray}
  \rho_{\rm GW}(t,k)
  = \frac{1}{32 \pi G}\frac{k^2}{a^2} \frac{k^3}{2 \pi^2}P_h(t,k),
\label{eq_GW_energy_h}
\end{eqnarray}
where $P_h$ is the power spectrum of GWs.  The GW spectrum
$\Omega_{\rm GW}(t,k)$ is defined as
\begin{eqnarray}
\Omega_{\rm GW}(t,k)
\equiv \frac{\rho_{\rm GW}(t,k)}{\rho_{{\rm tot}}(t)}.
\end{eqnarray}
For the GW modes entering the horizon at the RD era, the present value
of $\Omega_{\rm GW}$ is evaluated as
\begin{eqnarray}
  \Omega_{\rm GW}(t_0,k)
  \simeq 7.9 \times 10^{-15}
  \left( \frac{g_*(T_{\rm hi})}{g_*(T_{\rm eq})^{\rm (std)}} \right)
  \left( \frac{g_{*s}(T_{\rm eq})}{g_{*s}(T_{\rm hi})} \right)^{4/3}
  \left( \frac{k}{k_0} \right)^{n_t} r,
  \label{O_GW_t0}
\end{eqnarray}
where $r$ is the tensor-to-scalar ratio, $T_{\rm eq}$ is the temperature at the matter-radiation equality.\footnote{
	\label{footnote:3}
	Since at least two species of neutrinos have masses and non-relativistic at present, we use $g_*$ 
	at $T=T_{\rm eq}$ rather than that at $T=T_0$ (present temperature) to avoid confusion.
	At $T=T_{\rm eq}$, all neutrinos are relativistic. See also Appendix \ref{app:tensor}.
}
(Here, we assume that there is no entropy production after the horizon entry.)  $\Omega_{\rm GW}$
is weakly dependent on $k$ through $T_{\rm hi}(k)$ and the tensor
spectral index $n_t$.  The relation between the present frequency of
GWs and the temperature of the universe at which the corresponding
mode entered the horizon is given by
\begin{equation}
  f = \frac{k}{2\pi} \simeq 3.0\,{\rm Hz}
  \left( \frac{T}{10^8\,{\rm GeV}} \right)
  \left( \frac{g_*(T)}{228.75} \right)^{1/6}.
\end{equation}
Since future space-based GW detectors are most sensitive to the GW
with frequency around $0.1$--$1$\,Hz, we can probe the early universe
physics with very high temperature through the GW observations.

\subsubsection{GW spectrum: modes entering the horizon at the non-RD era}

In the above, we have assumed the RD universe at the horizon entry
of GWs.  This is not guaranteed in general in the early universe where
various forms of fluid can dominate the energy density.  Let us
suppose that the universe has the equation of state $w (> -1/3)$, the
ratio of the energy density to the pressure, at the horizon entry
of GWs.  Then the GW spectrum scales as
\begin{equation}
	\Omega_{\rm GW}(t_0,k) \propto k^{\frac{2(3w-1)}{3w+1}}.
\end{equation}
In the MD universe $(w=0)$, $\Omega_{\rm GW}(t_0,k) \propto k^{-2}$.

If there is a short period of inflation, the expression is a bit complicated.
Let us suppose that the equation of state changes as $w_1 \to w_2 \to w_3$ with $w_1,w_3 > -1/3$ and $w_2 < -1/3$.
The calculation is done in Appendix~\ref{sec:GW_inf} and the result is
\begin{equation}
	\Omega_{\rm GW}(t_0,k) \propto k^{2-\frac{4}{3w_1+1}+\frac{4}{3w_2+1}-\frac{4}{3w_3+1}  }.
\end{equation}
for the modes entering the horizon at the intermediate regime.  If
$w_1=w_3=1/3$ and $w_2=-1$, we obtain $\Omega_{\rm GW}(t_0,k) \propto
k^{-4}$.  If $w_1=1/3, w_3=0$ and $w_2=-1$, we obtain $\Omega_{\rm
  GW}(t_0,k) \propto k^{-6}$.

\subsection{Evolution of GWs: effects of dark radiation}

\subsubsection{Evolution equation}

In the presence of relativistic particles with weak or no interaction,
the RHS of \eqref[eq_GW_eom_u] does not vanish and affects the
evolution of
GWs~\cite{Weinberg:2003ur,Dicus:2005rh,Watanabe:2006qe,Ichiki:2006rn,Jinno:2012xb}.
Let us consider ``dark radiation'' $X$, non-interacting relativistic
particles, contributing to the anisotropic stress.\footnote
{We call non-interacting relativistic degrees of freedom at the time
  of our interest as ``dark radiation.''  Thus, the dark radiation in our
discussion may be massive, and may not correspond to the dark radiation
in the present universe.}
The RHS of
\eqref[eq_GW_eom_u] is written as an integration including the metric
perturbation $h(u,\mb[k],\lambda)$ (see Ref.~\cite{Jinno:2012xb} for
derivation),
\begin{eqnarray}
h^{''}(u) + 2 H_u h^{'}(u) + h(u)
= -24 \left[ H_{u}^2 \frac{1}{a^4 \rho_{\rm tot}} \right](u) \int_0^{u} du' \left[ a^4 \rho_X \pardif[h][u] \right](u') \frac{j_2 (u - u')}{(u - u')^2},
\label{eq_GW_eom_aniso}
\end{eqnarray}
where $\rho_X$ is the energy density of $X$, and $j_2(u)=\left[
  (3-u^2)\sin(u) - 3u\cos(u) \right] / u^3$.  Also, we have omitted
trivial indices.  The right-hand side of \eqref[eq_GW_eom_aniso] is
due to the backreaction of dark radiation on GWs with kernel
$j_2(u)/u^2$.  Note that $j_2(u)/u^2$ is suppressed when $u\gg 1$
while $h'(u) \simeq 0$ at $u \ll 1$, hence the anisotropic stress
affects the evolution of GWs only around $u \sim 1$, i.e., around the
horizon entry in the RD or MD universe.

\subsubsection{Overall normalization}

The presence of non-interacting particle $X$ at the time of horizon
entry or the last-scattering causes a change in the overall
normalization of the GW spectrum as shown in Ref.~\cite{Jinno:2012xb}.
Here we briefly repeat the result of Ref.~\cite{Jinno:2012xb}.

First, the amount of dark radiation is written in terms of the effective neutrino species $N_{\rm eff}$ as
\begin{eqnarray}
\rho_\nu + \rho_X
= N_{\rm eff} \frac{7}{8} \left( \frac{4}{11} \right)^{4/3} \rho_\gamma,
\end{eqnarray}
where $\rho_\nu$ and $\rho_\gamma$ are the energy densities of the
neutrino and photon, respectively.  With a standard matter content one
expects $N_{\rm eff}^{\rm (SM)} = 3.046$, and we define
\begin{eqnarray}
\Delta N_{\rm eff}
= N_{\rm eff} - N_{\rm eff}^{\rm (SM)}.
\end{eqnarray}
The dependence of $\Omega_{\rm GW}(t_0,k)$ on $\Delta N_{\rm eff}$
comes from $g_{*}$ and $g_{*s}$ in \eqref[O_GW_t0],
\begin{eqnarray}
\Omega_{\rm GW}(t_0,k)
\propto  \Omega_{{\rm r}, 0} \gamma \propto g_{*}(T_0) \gamma, ~~~
\gamma\equiv \left( \frac{g_{\ast}(T_{\rm hi})}{g_{\ast}(T_{\rm eq})} \right) \left( \frac{g_{\ast s}(T_{\rm eq})}{g_{\ast s}(T_{\rm hi})} \right)^{4/3}.
\end{eqnarray}
Then, we obtain
\begin{eqnarray}
\gamma
= \frac{1+\frac{7}{43} \left( \frac{g_{*s}(T_{\rm hi})}{10.75} \right)^{1/3} \Delta N_{\rm eff}}
{1/\gamma^{\rm (std)}+\frac{7}{43} \left( \frac{g_{*s}(T_{\rm hi})}{10.75} \right)^{1/3} \Delta N_{\rm eff}},
\end{eqnarray}
where, here and hereafter, the superscript ``(std)'' is for quantities 
in the case without dark radiation (i.e., $\Delta N_{\rm eff}=0$). 
Also, the relation between $g_{*}(T_{\rm eq})$ and $g_{*}^{\rm (std)}(T_{\rm eq})$ is
\begin{eqnarray}
\frac{g_{*}(T_{\rm eq})}{g_{*}^{\rm (std)}(T_{\rm eq})}
= \frac{2\left[ 1+N_{\rm eff} \cdot \frac{7}{8} \cdot \left( \frac{4}{11} \right)^{4/3} \right]}
{2\left[ 1+N_{\rm eff}^{\rm (std)} \cdot \frac{7}{8} \cdot \left( \frac{4}{11} \right)^{4/3} \right]}.
\end{eqnarray}
Then we define the overall factor $C_1$ as
\begin{eqnarray}
C_1\equiv  \frac{g_{*}(T_{\rm eq})}{g_{*}^{\rm (std)}(T_{\rm eq})} \frac{\gamma}{\gamma^{\rm(std)}},
\end{eqnarray}
which is plotted in \figref[fig_deltan_C]. 

\begin{figure}
  \centerline{\epsfxsize=\textwidth \epsfbox{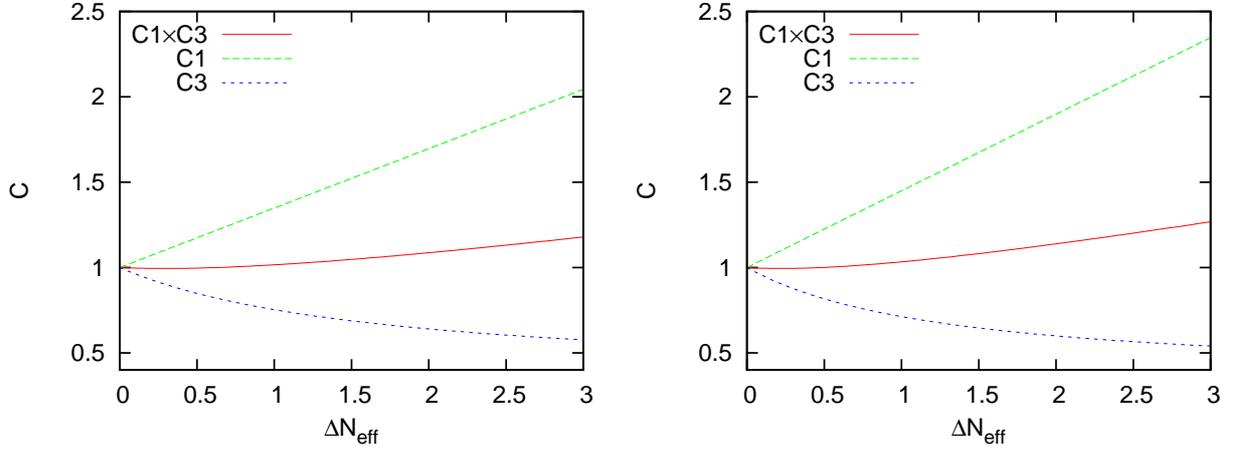}}
  \caption{\small Normalization factor $C$ with the presence of dark
    radiation.  $C_1$ accounts for the effect of increase in the
    amount of total radiation, while $C_3$ gives that of anisotropic
    stress due to the dark radiation. The left figure is for the SM,
    while the right is for the MSSM.  }
  \label{fig_deltan_C}
\end{figure}
\begin{figure}
  \centerline{\epsfxsize=\textwidth \epsfbox{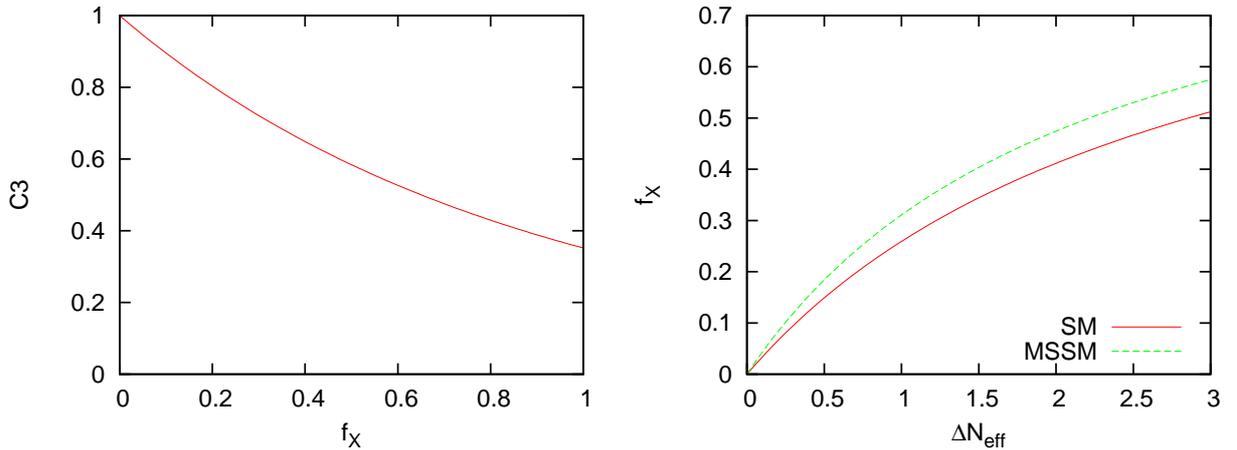}}
  \caption{\small Left : Normalization factor $C_3$ as a function of
    $f_X$, the energy fraction of dark radiation.  Right : Relation
    between the effective neutrino number $\Delta N_{\rm eff}$ and
    $f_X$.}
  \label{fig_deltan_f_C3}
\end{figure}

This $C_1$ factor is partially compensated by the effect of
anisotropic stress caused by $X$.  If $X$ exists before the horizon
entry, the RHS of \eqref[eq_GW_eom_aniso] can be simplified
as~\cite{Weinberg:2003ur}
\begin{eqnarray}
{\rm RHS \; of \; \eqref[eq_GW_eom_aniso] \;}
= -24 [H_u^2 f_X](u) \int_0^{u} du' \pardif[h][u] (u') \frac{j_2 (u - u')}{(u - u')^2},
\end{eqnarray}
where $f_X = \rho_X/\rho_{\rm tot}$ is the energy fraction of $X$.  In
this case, assuming the RD universe, the suppression of GW spectrum
caused by $X$ is calculated analytically.  The suppression factor is
defined as\footnote{The notations $C_1$ and $C_3$ follow those
    of Ref.~\cite{Boyle:2005se}. }
\begin{eqnarray}
C_3 
\equiv \frac{\Omega_{\rm GW} \left.\right|_{\rm w/ \; stress}}{\Omega_{\rm GW} \left.\right|_{\rm w/o \; stress}},
\end{eqnarray}
whose dependence on $f_X$ was analytically derived in
Refs.~\cite{Dicus:2005rh,Boyle:2005se} and plotted in
\figref[fig_deltan_f_C3].  The relation between $f_X$ and $\Delta
N_{\rm eff}$ is given by
\begin{eqnarray}
f_X
= \frac{\frac{7}{4} \left( \frac{4}{43} \right)^{4/3} \left[ g_{*s}^{\rm (std)}(T_{\rm hi}) \right]^{3/4} \Delta N_{\rm eff}}
{g_{*}^{\rm (std)}(T_{\rm hi}) + \frac{7}{4} \left( \frac{4}{43} \right)^{4/3} \left[ g_{*s}^{\rm (std)}(T_{\rm hi}) \right]^{3/4} \Delta N_{\rm eff}}.
\end{eqnarray}
Then the present GW spectrum reads
\begin{equation}
  \Omega_{\rm GW}(t_0,k) = C_1 C_3 \times \Omega_{\rm GW}^{\rm (std)}(t_0,k)
\end{equation}
for the mode entering the horizon at the RD era.  This gives the
overall normalization of the GW spectrum in the low-frequency limit,
$k < k_{\rm EW}$, where $k_{\rm EW}$ is the comoving Hubble scale at
around the electroweak phase transition.  The total modification on
the overall normalization on $\Omega_{\rm GW}$, $C_1 \times C_3$, is
plotted in \figref[fig_deltan_C].

In the following sections, we calculate the GW spectrum by using 
numerical calculation, taking account of the effects of anisotropic 
stress.  However, the normalization related to the $C_1$ factor is 
not included in the following calculations because it is 
model-dependent; $C_1$ depends whether $X$ remains non-interacting 
and relativistic until present or not.  Thus one should note that 
$\Omega_{\rm GW}$ given in the figures in the following sections 
should be multiplied by $C_1$ if the $X$ particle behaves as dark 
radiation until today.

\section{Illustration with simple examples}
\label{sec_Illustration}
\setcounter{equation}{0}

As we have mentioned, the spectrum of the inflationary GWs is
sensitive to the thermal history of the universe.  To see basic
features of GW spectrum, in this section, we exhibit some simple
examples for the GW spectrum modified by the phase transition, entropy
production and dark radiation.

\subsection{Phase transition} \label{sec:PT}

As a first example, let us consider a scalar field $\phi$ with a
symmetry breaking potential, e.g., $V = \lambda(\phi^2-v^2)^2$.  We
consider the case where $\phi$ is trapped at the origin due to thermal
effects \cite{Jinno:2011sw}.  If the interaction of $\phi$ with the
particles in thermal bath is strong enough, the vacuum energy
dominates the universe, and a brief period of inflation occurs, as in
the case of thermal inflation~\cite{Yamamoto:1985rd,Lyth:1995hj}.  The
phase transition happens after inflation and the subsequent
oscillation of the field is assumed to instantly decay into radiation.
The background equations we solve are \eqref[eq_Friedmann] with
\begin{eqnarray}
&&\rho_{\rm vac} =
\begin{cases}
\Lambda^4 \;\; &(t<t_{\rm PT}) \\
0 \;\; &(t>t_{\rm PT}) \\
\end{cases}
, \\
&&\dot{\rho_{\rm r}} + 4H\rho_{\rm r} = \Lambda^4 \delta(t-t_{\rm PT}),
\end{eqnarray}
where $t_{\rm PT}$ is the cosmic time at the phase transition.  We
numerically solved these with \eqref[eq_GW_eom_u], and the resultant
$\Omega_{\rm GW}$ is shown in \figref[fig_PT_RD_twin].  We varied the
ratio of $\rho_{\rm r}$ to $\rho_{\rm tot} \equiv \rho_{\rm r} +
\rho_{\rm vac}$ at the time of phase transition. The horizontal axis
is normalized with $k_{\rm PT}$, which satisfies $k=aH$ at the phase
transition.  Note that the ratio of the spectrum in $k \gg k_{\rm PT}$
to that of $k \ll k_{\rm PT}$ is equal to $\rho_{\rm r}/\rho_{\rm
  tot}$ just before the phase transition, which is from the fact that
$\rho_{\rm GW}(k\gg aH) \propto a^{-4}$.  GWs inside the horizon at
the phase transition are diluted by the newly-produced radiation,
while those outside the horizon remains $h_{ij}=$ const.  Note the
characteristic oscillatory feature around $k \simeq k_{\rm PT}$.  The
reason for the oscillatory feature and the oscillation period is
explained in Appendix \ref{sec:GW_inf}.  Finally, note that in the
limit of long duration of thermal inflation, the spectrum scales as
$\propto k^{-4}$ as shown also in Appendix \ref{sec:GW_inf}.

\begin{figure}
  \centerline{\epsfxsize=\textwidth\epsfbox{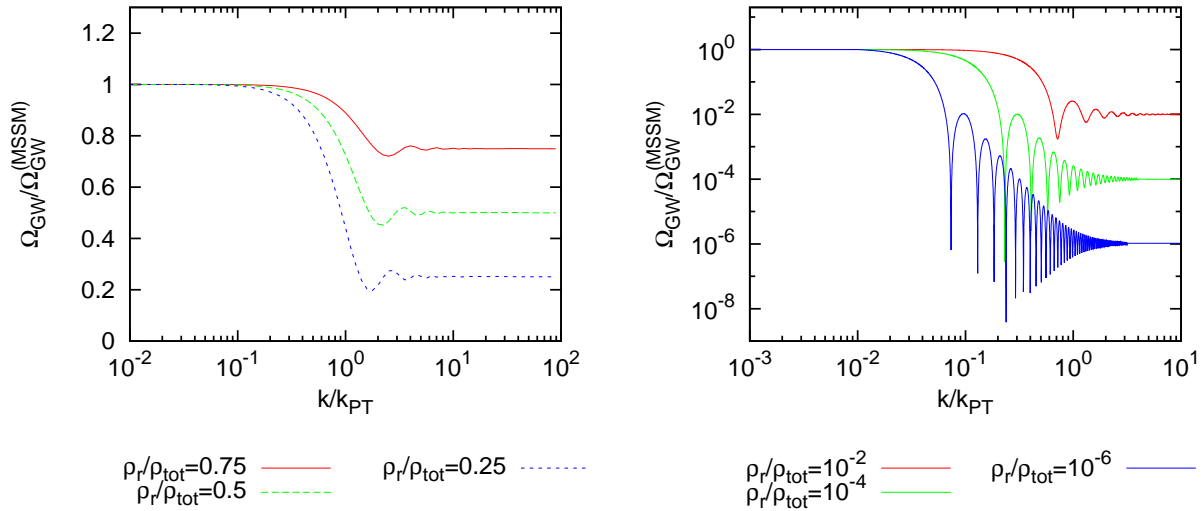}}
  \caption{\small GW spectrum with phase transition and instant decay
    into radiation.  We assumed that the universe is
    radiation dominated before the vacuum energy dominates it, and
    varied the ratio of radiation energy density to the total energy
    density at the phase transition. }
  \label{fig_PT_RD_twin}
\end{figure}

\subsection{Entropy production}  \label{sec:EP}

\begin{figure}
  \centerline{\epsfxsize=\textwidth\epsfbox{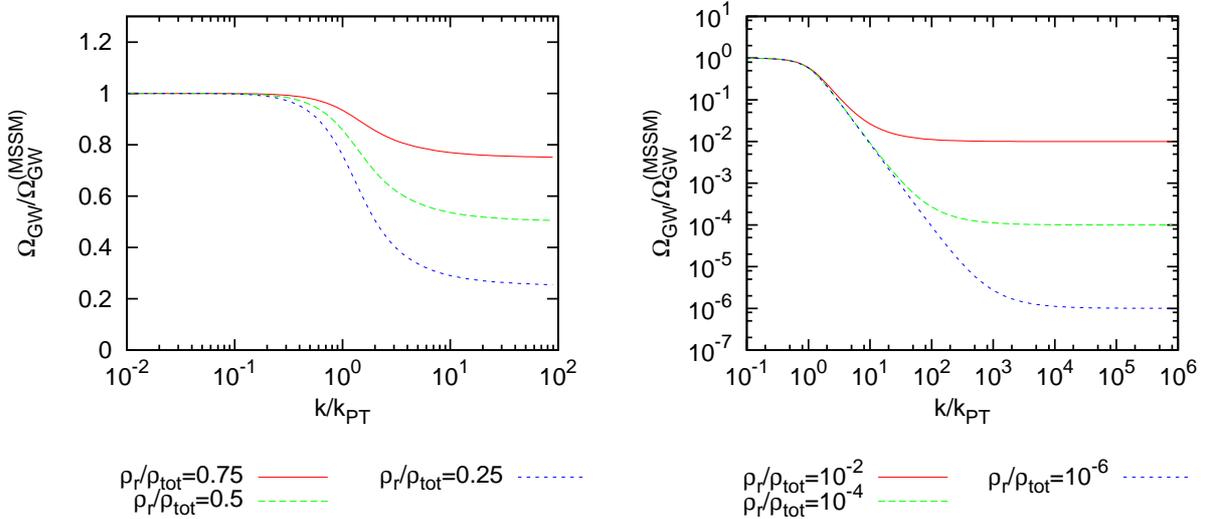}}
  \caption{\small GW spectrum with entropy injection. We varied the ratio of 
  radiation energy density to the total energy density at $t=t_{\rm dec}\equiv 
  \Gamma^{-1}$.  
  }
  \label{fig_EI_twin}
\end{figure}

Next we consider the case of late-time entropy production, i.e., the
case where some matter dominates universe, and then it decays into
radiation.  The background equations we solve are \eqref[eq_Friedmann]
and
\begin{eqnarray}
&&\dot{\rho}_{\rm m} +3H\rho_{\rm m} = -\Gamma \rho_{\rm m}, \\
&&\dot{\rho}_{\rm r}  + 4H\rho_{\rm r} = \Gamma \rho_{\rm m},
\end{eqnarray}
where $\Gamma$ is the decay rate of the non-relativistic matter.  The
result is shown in \figref[fig_EI_twin].  We varied the ratio of
$\rho_{\rm r}$ to $\rho_{\rm tot} \equiv \rho_{\rm r} + \rho_{\rm m}$
at the decay.  The horizontal axis is normalized with $k_{\rm decay}$,
which is defined as the comoving Hubble scale $aH$ at $t=\Gamma^{-1}$.
Note that the ratio of the spectrum in $k \gg k_{\rm decay}$ to that
of $k \ll k_{\rm decay}$ is equal to $\rho_{\rm r}/\rho_{\rm tot}$ at
the decay for the same reason written in the previous subsection.  In
this case the spectrum scales as $\propto k^{-2}$ for the mode
entering the horizon at the non-relativistic matter dominated era.
Also note that there is no oscillatory feature in the GW spectrum in
contrast to the previous case.



\subsection{Phase transition and dark radiation}

Let us consider the case where vacuum energy of a scalar field
dominates the universe as in the case of Sec.~\ref{sec:PT}, but at a
certain time the energy is instantly converted to dark radiation $X$.
We numerically solved the Friedmann equation \eqref[eq_Friedmann] with
\begin{eqnarray}
&&\rho_{\rm vac} =
\begin{cases}
\Lambda^4 \;\; &(t<t_{\rm PT}) \\
0 \;\; &(t>t_{\rm PT}) \\
\end{cases}
, \\
&&\dot{\rho}_r + 4H\rho_{\rm r} = 0, \\
&&\dot{\rho_X} + 4H\rho_X = \Lambda^4 \delta(t-t_{\rm PT}).
\end{eqnarray}
Then we expect the following features in the GW spectrum:
\begin{itemize}
\item For large wavenumber $k \gg k_{\rm PT}$, there is the same
  suppression as \figref[fig_PT_RD_twin].
\item For small wavenumber $k \ll k_{\rm PT}$, there is a suppression
  due to the anisotropic stress, caused by the RHS of
  \eqref[eq_GW_eom_aniso].
\end{itemize}
The reason for no suppression by the anisotropic stress at $k \gg
k_{\rm PT}$ is that, for GWs of such large wavenumber, there do not
exist $X$ particles at the time of their horizon entry.  The results
of numerical calculations are shown in \figref[fig_aniso_PT1].  In the
figure we varied $\Lambda$ so that $\Delta N_{\rm eff}$ (after the
phase transition) becomes $1, 2, 5$ and $100$.  Note that $\Delta
N_{\rm eff}$ here is evaluated assuming that $X$ is relativistic
and survives until today.  If $X$ decays into radiation at some
epoch, or if there exists another entropy production, the $X$
abundance at the epochs of the BBN and radiation-matter equality can
be reduced, hence $\Delta N_{\rm eff} \gg 1$ does not necessarily
conflict with observations.  In such a case the overall normalization
of the GW spectrum changes, but the shape of the spectrum does not
change.  One sees a dip around $k \simeq k_{\rm PT}$, which could be a
smoking-gun signal of the phase transition followed by
the production of dark radiation.
    
\begin{figure}
  \begin{minipage}{0.5\columnwidth}
    \begin{center}
      \includegraphics[clip, width=1.0\columnwidth]{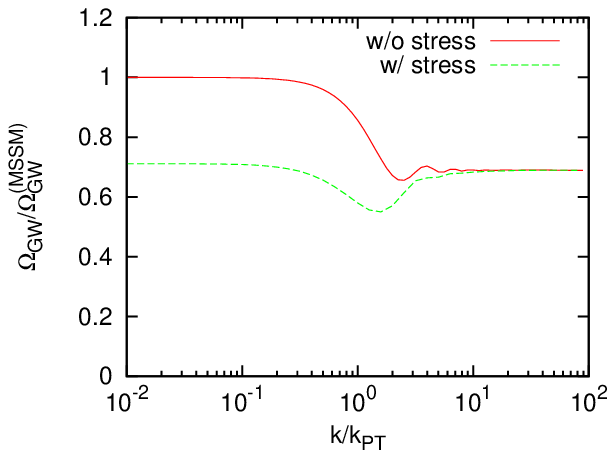}
    \end{center}
  \end{minipage}%
  \begin{minipage}{0.5\columnwidth}
    \begin{center}
      \includegraphics[clip, width=1.0\columnwidth]{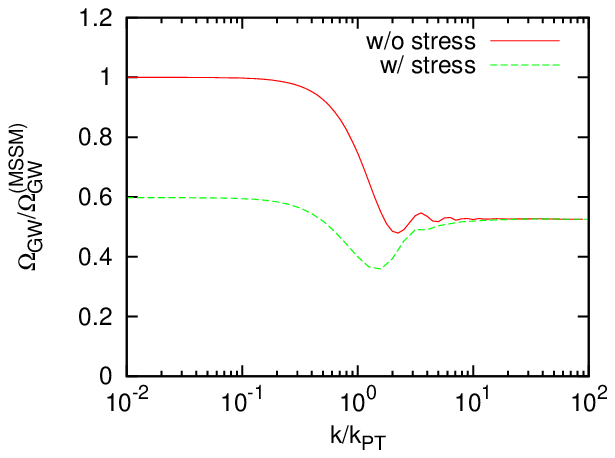}
    \end{center}
  \end{minipage}
  
  \begin{minipage}{0.5\columnwidth}
    \begin{center}
      \includegraphics[clip, width=1.0\columnwidth]{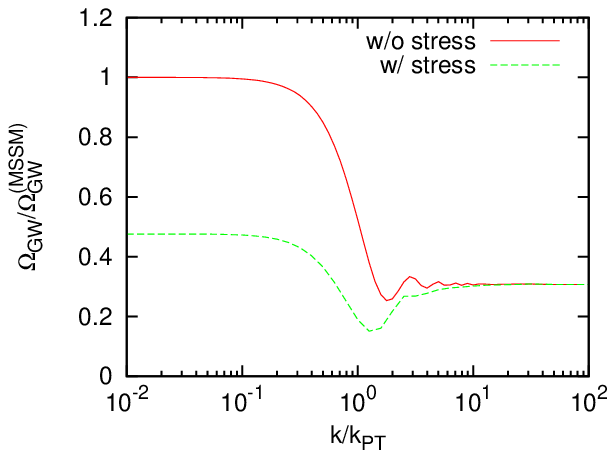}
    \end{center}
  \end{minipage}%
  \begin{minipage}{0.5\columnwidth}
    \begin{center}
      \includegraphics[clip, width=1.0\columnwidth]{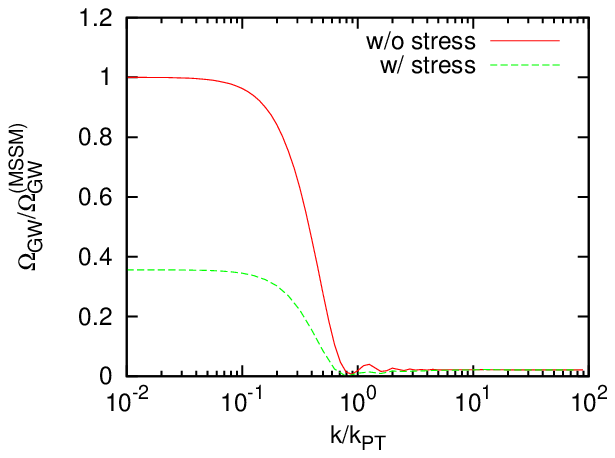}
    \end{center}
  \end{minipage}
  
  \caption{\small GW spectrum with phase transition and subsequent
    instant decay into dark radiation $X$.  We have assumed that the
    branching ratio to $X$ is 0 and 1 in the red-solid and
    green-dashed line, respectively.  Also, $\Delta N_{\rm eff}=1, 2,
    5$ and $100$ for the top left, top right, bottom left and bottom
    right figure, respectively. }
  \label{fig_aniso_PT1}
\end{figure}

\subsection{Entropy production and dark radiation}

Next let us consider the case where some massive particle dominates
the universe as in the case of Sec.~\ref{sec:EP}, but then it decays
into dark radiation $X$ \cite{Jinno:2012xb}. We numerically solved
\eqref[eq_Friedmann] and
\begin{eqnarray}
&&\dot{\rho}_{\rm m} +3H\rho_{\rm m} = - \Gamma \rho_{\rm m}, \\
&&\dot{\rho}_{\rm r} + 4H\rho_{\rm r} = 0, \\
&&\dot{\rho}_X + 4H\rho_X = \Gamma \rho_{\rm m}.
\end{eqnarray}
The results of numerical calculations are shown in \figref[fig_aniso_EI1]. 
One also sees a characteristic dip around $k \simeq k_{\rm PT}$ as in the previous case.

\begin{figure}
  \begin{minipage}{0.5\columnwidth}
    \begin{center}
      \includegraphics[clip, width=1.0\columnwidth]{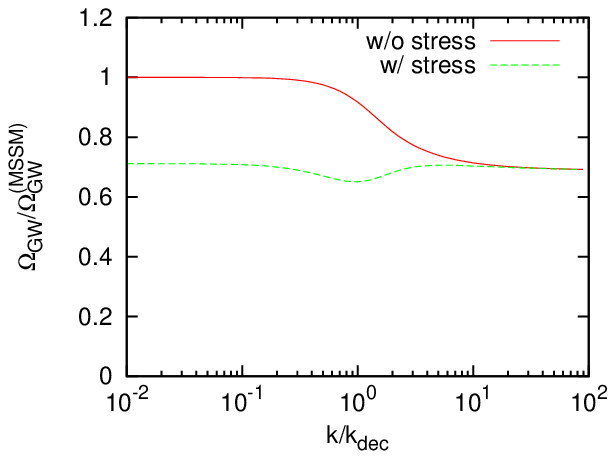}
    \end{center}
  \end{minipage}%
  \begin{minipage}{0.5\columnwidth}
    \begin{center}
      \includegraphics[clip, width=1.0\columnwidth]{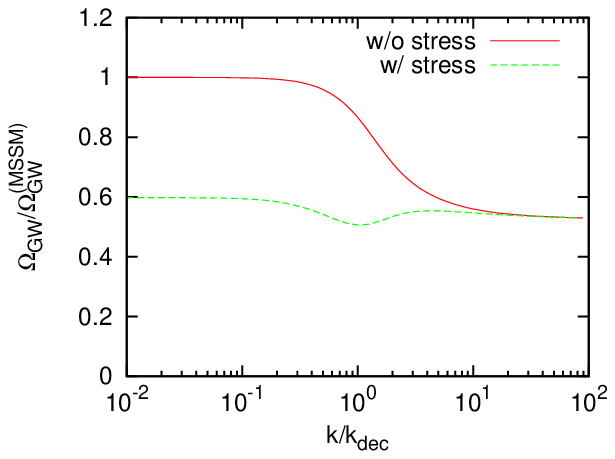}
    \end{center}
  \end{minipage}
  
  \begin{minipage}{0.5\columnwidth}
    \begin{center}
      \includegraphics[clip, width=1.0\columnwidth]{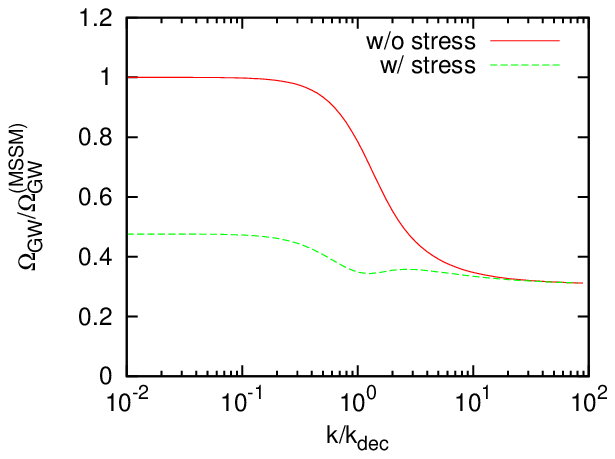}
    \end{center}
  \end{minipage}%
  \begin{minipage}{0.5\columnwidth}
    \begin{center}
      \includegraphics[clip, width=1.0\columnwidth]{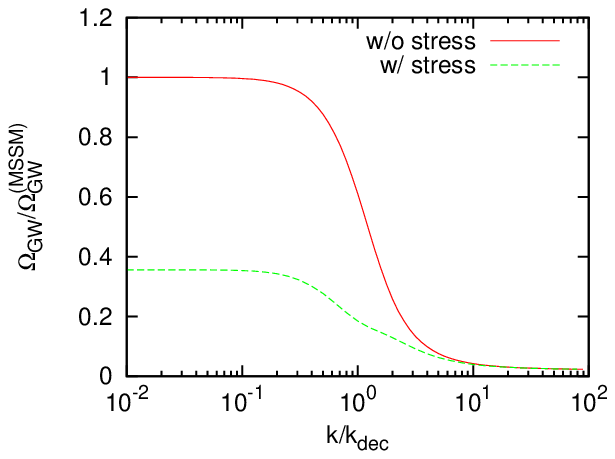}
    \end{center}
  \end{minipage}
  
  \caption{\small GW spectrum with the decay of massive particle
    into dark radiation.  The horizontal axis is normalized by
    $k_{\rm dec}\equiv aH(t=\Gamma^{-1})$, We have assumed that
    the branching ratio to $X$ is 0 and 1, in the red-solid and
    green-dashed line, respectively. Also, $\Delta N_{\rm
      eff}=1, 2, 5$ and $100$ for the top left, top right,
    bottom left and bottom right figure, respectively. }
  \label{fig_aniso_EI1}
\end{figure}

\subsection{Decay of dark radiation into visible radiation}

Another interesting possibility is that $X$ is produced at some time or has existed from the beginning, and 
then decays into visible radiation while $X$ is still relativistic. 
In this case the background equations we should solve are
\begin{eqnarray}
&&\pardif[\ln (a^3 F_X)][t]
= H\pardif[\ln (a^3 F_X)][\ln E] - \frac{m \Gamma}{E},
\label{FXdot}
\\
&&\dot{\rho}_X + \dot{\rho}_{\rm r} + 4H(\rho_X + \rho_{\rm r} )
= 0,
\label{rhoXdot_dr2vs}
\end{eqnarray}
where $m$ is the (small) mass of $X$, $\Gamma$ is the decay rate of
$X$ and $F_X (t,E)$ is defined so that $X$ with energy $E$ -- $E+dE$
carries energy density of $F_X (t,E) dE$ at the time $t$.\footnote
{Without decay, the energy density carried by $X$ with energy $E$ --
  $E+dE$ in a comoving volume falls proportional to $a^{-1}$:
  \begin{eqnarray*}
    a^3(t+dt) F_X(t+dt,E(t+dt)) dE(t+dt)
    = \frac{a(t)}{a(t+dt)} a^3(t) F_X(t,E(t)) dE(t).
  \end{eqnarray*}
  Since we know that $E(t)$, $dE(t) \propto a^{-1}$, we obtain
  \begin{eqnarray*}
    \pardif[\ln (a^3 F_X)][t] - H\pardif[\ln (a^3 F_X)][\ln E] = 0.
  \end{eqnarray*}
  If we include decay, the energy density decreases with decay rate
  $\Gamma$ suppressed by the $\gamma$-factor:
  \begin{eqnarray*}
    a^3(t+dt) F_X(t+dt,E(t+dt)) dE(t+dt) 
    &=& \frac{a(t)}{a(t+dt)} a^3(t) F_X(t,E(t)) dE(t) 
    \nonumber \\ &&    
    - \frac{\Gamma}{E(t)/m} a^3(t) F_X(t,E(t)) dE(t).
  \end{eqnarray*}
  Again using $E(t)$, $dE(t) \propto a^{-1}$, we get
  \eqref[FXdot].}
It satisfies $\int dE F_X(t,E) = \rho_X(t)$.  We solved
\eqref[eq_Friedmann], \eqref[eq_GW_eom_aniso], \eqref[FXdot], and
\eqref[rhoXdot_dr2vs], varying the initial ratio of $\rho_X$ to
$\rho_{\rm r}$ and the energy dependence of $F_X$.  The result is
shown in \figref[fig_rel_decay] for the case where the dark radiation
initially dominates the universe.  We consider three cases with
different distribution functions: the line (i.e., $F_X (t,E) \propto
\delta(E_0)$), bosonic, and fermionic thermal distributions.  Note
that the spectrum has a hill around $k \simeq k_{\rm dec}$, where
$k_{\rm dec}$ for the line distribution is defined as the comoving
Hubble scale $k=aH$ at $t=(m \Gamma/E)^{-1}$. For thermal distribution
it is defined as $k=aH$ at $t=(m \Gamma/\tilde{E})^{-1}$ with
$\tilde{E} \simeq T$.  We found no significant differences in the GW
spectrum among these distributions.  Thus, in studying the
relativistic decay of dark radiation in the next section, we
approximate that it has the line spectrum.

\begin{figure}
  \centerline{\epsfxsize=9cm \epsfbox{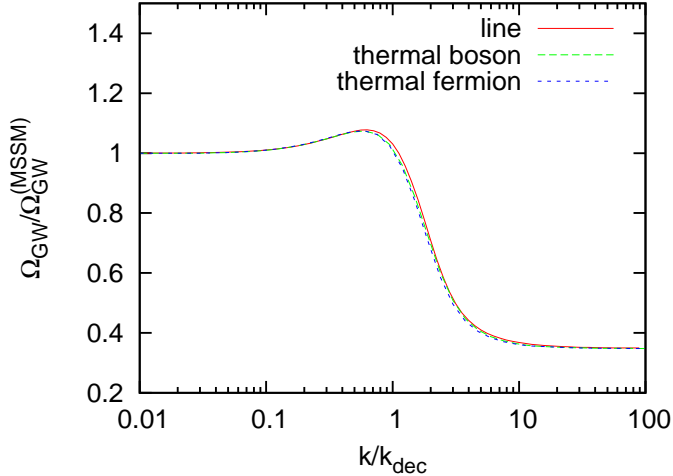}}
  \caption{\small GW spectrum with the decay of dark radiation $X$
    into visible radiation.  We varied the energy distribution of the
    dark radiation.  }
  \label{fig_rel_decay}
\end{figure}

\section{Examples of the GW spectrum}
\label{sec_Examples_of_possible_GW_spectrum}
\setcounter{equation}{0}

\begin{figure}
  \centerline{\epsfxsize=\textwidth \epsfbox{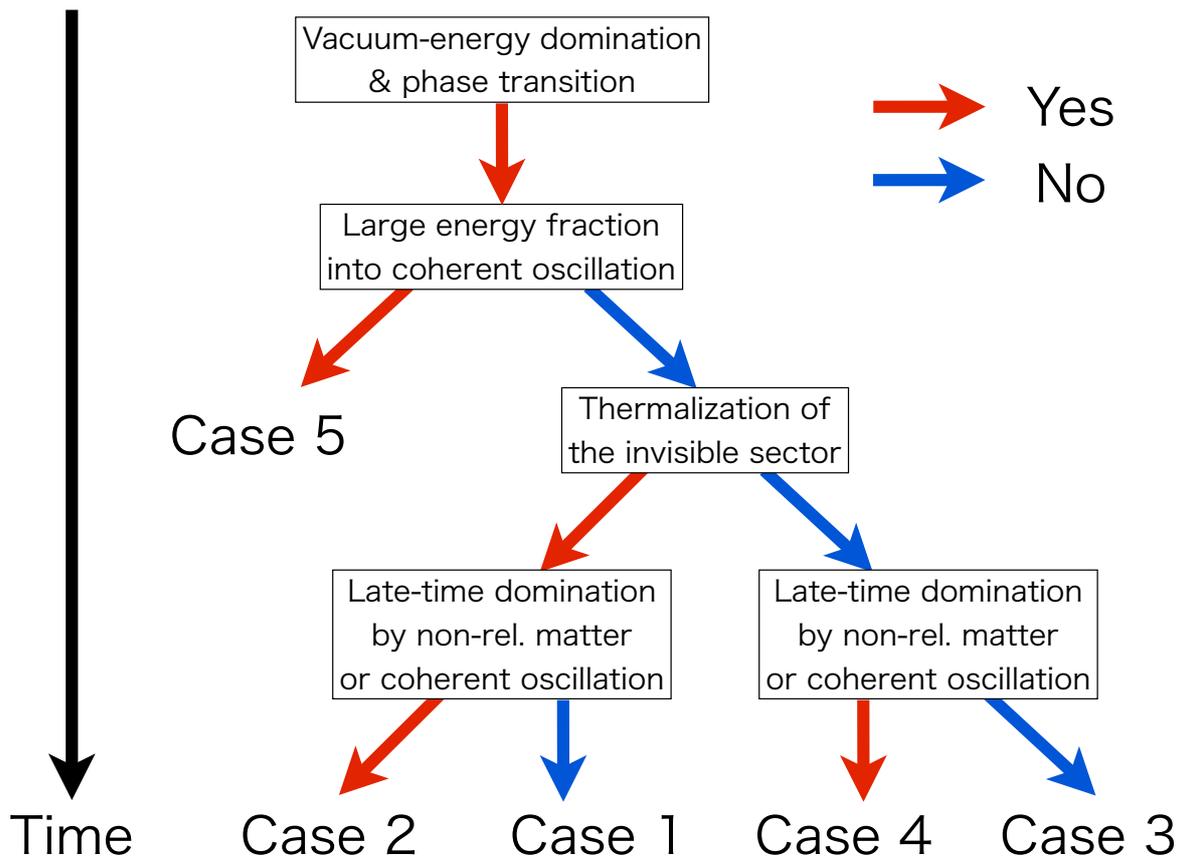}}
  \caption{\small The Cases discussed in this paper.}
  \label{fig:flowchart}
\end{figure}

In this section we study the spectrum of the GW in cases with the
combinations of events discussed in the previous section, which may be
realized in some models motivated by particle physics.  (See the flow
chart given in Fig.\ \ref{fig:flowchart}.)  In our study, we perform
the analysis as general as possible, without specifying underlying
models.  Examples of particle-physics models realizing the scenarios
in this section will be discussed in the next section.


\subsection{Case 1}   \label{sec:case1}

In this subsection we assume that the universe has undergone the
following events in a time ordering:
\begin{enumerate}
\item A brief period of thermal inflation is caused by a scalar field
  $\phi$.
\item After the phase transition, $\phi$ instantaneously decays into
  radiation with short mean free path.
\item $X$ particles decouple from the thermal bath, after which $X$
  particles behave as dark radiation.
\item Part of the decoupled particles $X$ decays into visible
  radiation.
\end{enumerate}
Before thermal inflation, the universe is assumed to be radiation
dominated.  Each component evolves as
\begin{eqnarray}
  &&\rho_{\rm vac} = \begin{cases}
    \Lambda^4 & ~~{\rm for}~~ t < t_{\rm PT}\\
    0                   & ~~{\rm for}~~ t > t_{\rm PT},
  \end{cases}
  \\
  &&\dot{\rho}_{\rm r} + 4H\rho_{\rm r} = 
  \Lambda^4\delta(t-t_{\rm PT})
  -\epsilon_X\rho_{\rm r}\delta(t-t_{\rm decouple})
  + \frac{m}{E}\Gamma \rho_{X_1},
  \\
  &&\dot{\rho}_{X_1} + \left(4H + \frac{m}{E}\Gamma \right) \rho_{X_1} = \epsilon_{X_1} \rho_{\rm r} \delta(t-t_{\rm decouple}),\\
  &&\dot{\rho}_{X_2} + 4H \rho_{X_2} = \epsilon_{X_2} \rho_{\rm r} \delta(t-t_{\rm decouple}),
  \label{eq_evolution_r}
\end{eqnarray}
where $\epsilon_X (= \epsilon_{X_1}+\epsilon_{X_2})$ is the fraction
of the radiation which becomes dark radiation $X$.  For simplicity,
the decoupling is assumed to occur instantaneously.  Dark radiation
$X$ is divided into two components $X_1$ and $X_2$, the former of
which decays into the radiation in the visible sector.  Both $X_1$ and
$X_2$ contribute to the anisotropic stress, RHS of
\eqref[eq_GW_eom_aniso].

The result of numerical calculation on the GW spectrum is shown in
\figref[fig_1_2_2].  Here, we consider the case where all the vacuum
energy eventually goes into $X$ particles which initially have short
mean free path.  Here, we used $\rho_{\rm r} / \rho_{\rm tot} = 0.63$
at the phase transition.  We have also taken
\begin{eqnarray}
T_{\rm decouple} &=& 10^{-2} \times T_{\rm PT}, \\
T_{\rm decay} &=& 10^{-4} \times T_{\rm PT},
\end{eqnarray}
to fix $t_{\rm decouple}$ and $t_{\rm decay}$, where $T_{\rm decay}$ is
defined as $H(T=T_{\rm decay}) = m \Gamma / T_{\rm decay}$.  Here $m$
is the mass of $X$ and $\Gamma$ is the decay rate of $X$ at rest.  We
have assumed $\epsilon_{X_1} = \epsilon_{X_2} = \epsilon_X/2$ and also
fixed $\epsilon_X$ so that $\Delta N_{\rm eff}$ becomes 0.5.
 
It is seen that the spectrum changes steeply at $k\simeq k_{\rm PT}$
due to the brief period of inflation.  Due to the presence of dark
radiation $X$, there is a suppression caused by the anisotropic stress
for $k \lesssim10^{-2} k_{\rm PT}$ because GWs with such wavenumber
enter the horizon after the decoupling of $X$.  Finally, since a part
of $X$ decays into visible radiation, the suppression becomes weaker
for GWs with $k \lesssim 10^{-4}k_{\rm PT}$ which enter the horizon
after the $X_1$ decay.

\begin{figure}
  \centerline{\epsfxsize=10cm\epsfbox{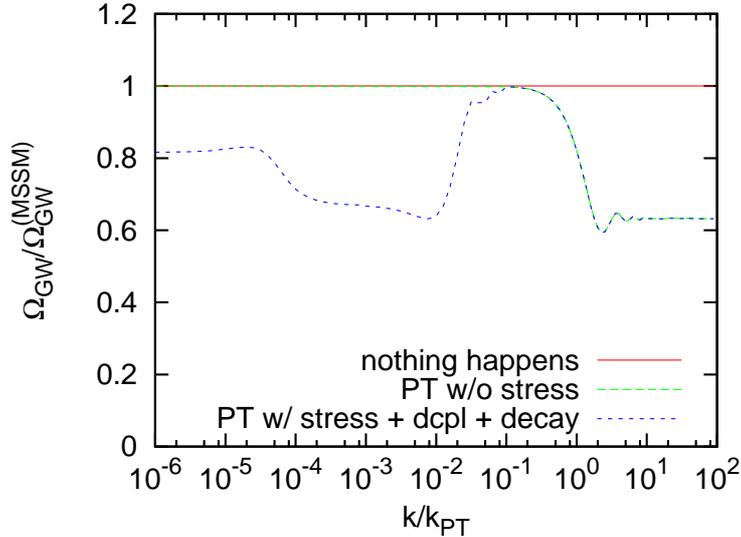}}
  \caption{\small (Blue-dotted) GW spectrum for Case 1, i.e, GW
    spectrum with decoupling of dark radiation $X$ and subsequent
    decay of $X$ in addition to the phase transition. $\Delta N_{\rm
      eff}$ is assumed to be 1.3 at the decoupling and then decreases
    to 0.5 after the decay.  (Red-solid) Flat spectrum expected in a
    simple RD universe.  (Green-dashed) GW spectrum for the case
    without the production of dark radiation.  }
  \label{fig_1_2_2}
\end{figure}

\subsection{Case 2}    \label{sec:case2}

Now, let us consider the Case 2, which has the following thermal
history.  (The first three events are the same as in the Case 1.)
\begin{enumerate}
\item A brief period of thermal inflation is caused by a scalar field
  $\phi$.
\item After the phase transition, $\phi$ instantaneously decays into
  radiation.
\item $X$ particles decouple from the thermal bath, after which $X$
  particles behave as dark radiation.
\item Some non-relativistic matter begins to dominate the universe.
\item The non-relativistic matter decays into the visible radiation.
\end{enumerate}
Part of visible radiation or $X$ may provide the non-relativistic
matter if it becomes non-relativistic due to the redshift.  In order
to study the case in which the universe evolves from the RD epoch to
the MD epoch, we adopt the following equations to follow the evolution
of the background:
\begin{eqnarray}
  &&\rho_{\rm vac} = \begin{cases}
    \Lambda^4 & ~~{\rm for}~~ t < t_{\rm PT}\\
    0                   & ~~{\rm for}~~ t > t_{\rm PT},
  \end{cases}
  \\
  &&\dot{\rho}_{\rm r} + 4H\rho_{\rm r} = \Lambda^4\delta(t-t_{\rm PT})-\epsilon_X\rho_{\rm r}\delta(t-t_{\rm decouple})
  + \Gamma \rho_{\rm m},
  \\
  &&\dot{\rho}_{X} + 4H \rho_{X} = \epsilon_{X} \rho_{\rm r} \delta(t-t_{\rm decouple}),\\
  &&\dot{\rho}_{\rm m} + 3H\rho_{\rm m} = - \Gamma \rho_{\rm m},
\end{eqnarray}
where $t_{\rm decouple}\ll \Gamma^{-1}$.  The resulting GW spectra are
shown in \figref[fig_2_1_3_2] and \figref[fig_2_2_3_2].  Here, we
assumed instant decoupling as in the Case 1 and
\begin{eqnarray}
T_{\rm decouple} &=& 10^{-2} \times T_{\rm PT}, \\
T_{\rm decay} &=& 10^{-5} \times T_{\rm PT}.
\end{eqnarray}
In addition, in our calculation, the massive particle which dominates
the universe is assumed to originate from the visible radiation. If it
is part of the non-interacting radiation we may not apply the
derivation of the wave equation in Ref.~\cite{Jinno:2012xb} and the
calculation would be very complicated.  However, from the fact that
massive particles do not generate anisotropic stress, the GW spectrum
in such a case is expected to be almost the same in the above figure.

\begin{figure}
  \centerline{\epsfxsize=9cm\epsfbox{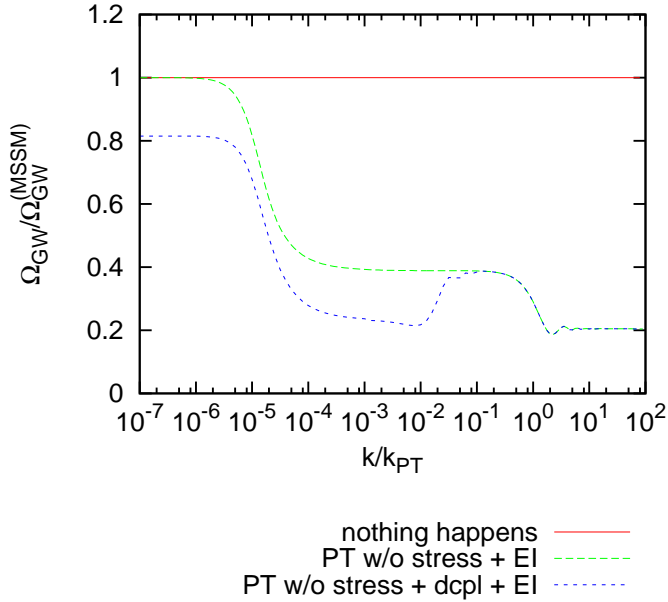}}
  \caption{\small (Blue-dotted) GW spectrum for Case 2; the case
    with the decoupling of some radiation from thermal bath.  $\Delta
    N_{\rm eff}$ is assumed to be 2 at the decoupling and then
    decrease to 0.5 after decay.  (Red-solid) Flat spectrum expected
    in simple RD universe.  (Green-dashed) Spectrum without the
    production of dark radiation.  }
   \label{fig_2_1_3_2}
\end{figure}
\begin{figure}
  \centerline{\epsfxsize=9cm\epsfbox{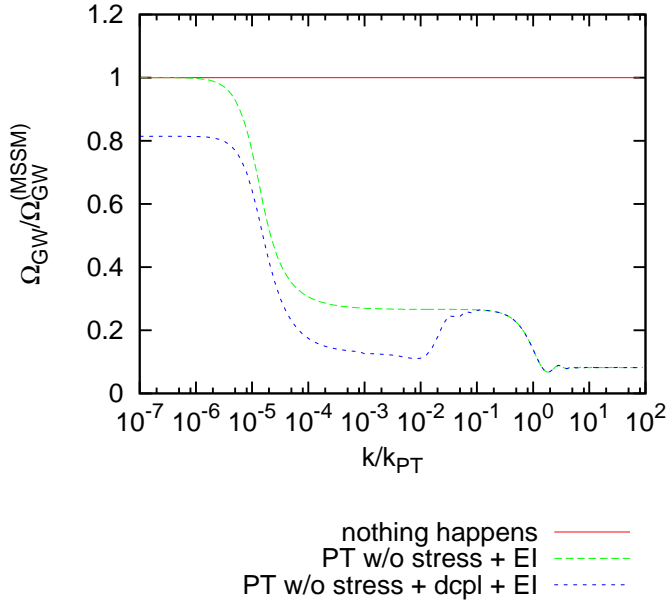}}
  \caption{\small GW spectrum for Case 2. Same as
    \figref[fig_2_1_3_2] except that $\Delta N_{\rm eff} = 5$ at the
    decoupling.}
  \label{fig_2_2_3_2}
\end{figure}


\subsection{Case 3}   \label{sec:case3}

Next, let us consider the following cosmological scenario.
\begin{enumerate}
\item A brief period of thermal inflation is caused by a scalar field
  $\phi$.
\item After the phase transition, $\phi$ instantaneously decays into
  dark radiation $X$.
\item Part of dark radiation decays into visible radiation.
\end{enumerate}
Here, the decay of $\phi$ is approximated to occur instantaneously so
that dark radiation has monochromatic spectrum (with energy $E$).
Then, each component evolves as
\begin{eqnarray}
  &&\rho_{\rm vac} = \begin{cases}
    \Lambda^4 & ~~{\rm for}~~ t < t_{\rm PT}\\
    0                   & ~~{\rm for}~~ t > t_{\rm PT},
  \end{cases}
  \\
  &&\dot{\rho}_{X_1} + 4H\rho_{X_1} = \epsilon_{X_1}\Lambda^4\delta(t-t_{\rm PT})-\frac{m\Gamma}{E} \rho_{\rm X_1},
  \\
  &&\dot{\rho}_{X_2} + 4H\rho_{X_2} = \epsilon_{X_2}\Lambda^4\delta(t-t_{\rm PT}),\\
  &&\dot{\rho}_{\rm r} + 4H \rho_{\rm r} = \frac{m\Gamma}{E} \rho_{X_1},
\end{eqnarray}
where $\epsilon_{X_1} + \epsilon_{X_2} = 1$.
A crucial difference from the previous two cases is that $\phi$ mainly decays into dark radiation
so that the effect of anisotropic stress is already significant just after the phase transition.

The numerical result on the GW spectrum is shown in \figref[fig_3_2].
We have assumed
\begin{eqnarray}
T_{\rm decay} &=& 10^{-2} \times T_{\rm PT}.
\end{eqnarray}
$\epsilon_{X_1}$ and $\epsilon_{X_2} $ are chosen so that
$\Delta N_{\rm eff}$ becomes 1.3 at the decoupling and 
then decreases to 0.5 after the decay $(\epsilon_{X_1}=\epsilon_{X_2}=1/2)$.
One finds a dip around $k \simeq k_{\rm PT}$ due to the anisotropic stress caused by $X$, 
as in \secref[sec_Illustration], instead of a hill seen in the previous cases.
In the low frequency limit $k \lesssim 10^{-2}k_{\rm PT}$, the suppression 
by the anisotropic stress is less efficient because $X_1$ does not exist 
when such modes enter the horizon.

\begin{figure}
  \centerline{\epsfxsize=9cm\epsfbox{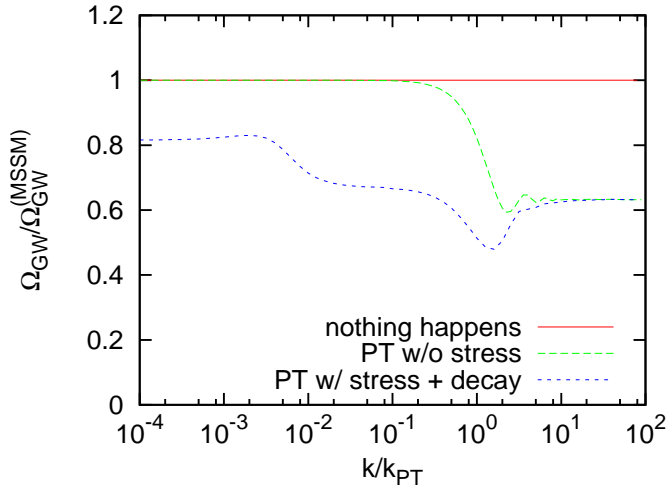}}
  \caption{\small (Blue-dotted) GW spectrum for Case 3; the case with
    a short period of inflation, phase transition, instant decay into
    dark radiation and the decay of the dark radiation.  $\Delta
    N_{\rm eff}$ is assumed to be 1.3 at the decoupling and then
    decrease to 0.5 after the decay.  (Red-solid) Flat GW spectrum
    expected in a simple RD universe.  (Green-dashed) GW spectrum for
    the case without dark radiation.  }
  \label{fig_3_2}
\end{figure}


\subsection{Case 4} \label{sec:case4}

Let us consider the following cosmological scenario with late-time
entropy production.  The first two are the same as the Case 3.
\begin{enumerate}
\item A brief period of thermal inflation is caused by a scalar field
  $\phi$.
\item After the phase transition, $\phi$ instantaneously decays into
  dark radiation $X$.
\item Non-relativistic matter dominates the universe.
\item Non-relativistic matter decays into radiation.
\end{enumerate}
Non-relativistic matter exists in many models of phase transition
since the coherent oscillation of the scalar field often survives
after the phase transition.  In this case, we use the following set of
evolution equations:
\begin{eqnarray}
  &&\rho_{\rm vac} = \begin{cases}
    \Lambda^4 & ~~{\rm for}~~ t < t_{\rm PT}\\
    0                   & ~~{\rm for}~~ t > t_{\rm PT},
  \end{cases}
  \\
  &&\dot{\rho}_{X} + 4H\rho_{X} = \Lambda^4\delta(t-t_{\rm PT}),
  \\
  &&\dot{\rho}_{\rm m} + 3H\rho_{\rm m} = -\Gamma\rho_{\rm m},\\
  &&\dot{\rho}_{\rm r} + 4H \rho_{\rm r} = \Gamma \rho_{\rm m},
\end{eqnarray}
where $t_{\rm PT}\ll\Gamma^{-1}$.

The resulting GW spectra are shown in \figref[fig_411_twin] and
\figref[fig_412_twin].  In these figures, we have varied the
parameters $R_{\rm rad}$ and $R_{\rm coh}$, which are defined as
\begin{eqnarray}
  R_{\rm rad} 
  \equiv
  \left.
    \frac{\rho_{\rm rad}}{\rho_{\rm tot}} 
  \right|_{\rm just \; before \; PT}
  = \left.
    \frac{\rho_{\rm rad}}{\rho_{\rm rad} + \rho_{\rm vac}}
  \right|_{\rm just \; before \; PT},
\end{eqnarray}
and 
\begin{eqnarray}
  R_{\rm coh} \equiv 
  \frac{\rho_{\rm coh}|_{\rm just \; after \; PT}}
  {\rho_{\rm vac}|_{\rm just \; before \;PT}},
  \label{R_coh}
\end{eqnarray}
with $\rho_{\rm coh}$ and $\rho_{\rm tot}$ being the energy density of
the coherent oscillation and the total energy density, respectively.
For the analysis of Case 4, we take $\rho_{\rm m}=\rho_{\rm coh}$.  We
have taken $R_{\rm coh}=0.1$ in \figref[fig_411_twin], and $R_{\rm
  coh}=0.01$ in \figref[fig_412_twin]. The value of $R_{\rm rad}$ is
chosen so that, with the assumption that the vacuum energy goes only
into the coherent oscillation and dark radiation, the effective
neutrino number $\Delta N_{\rm eff}$ at the phase transition and
subsequent reheating era is 2.  In each case $\Delta N_{\rm eff}$ is
diluted to 0.5 after the entropy production.  Note that the features
of phase transition with anisotropic stress and entropy injection
appear in the figure.

\begin{figure}
  \centerline{\epsfxsize=\textwidth\epsfbox{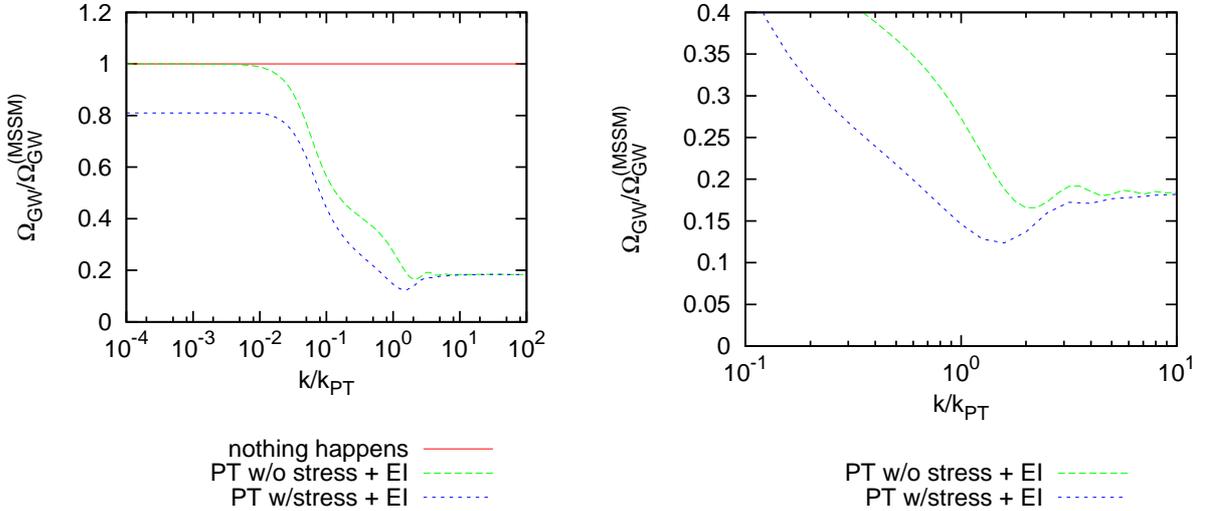}}
  \caption{\small Left: (Blue-dotted) GW spectrum for Case 4.  $R_{\rm
      coh}=0.1$, while $\Delta N_{\rm eff}=2$ and $0.5$ just after the
    phase transition and after decay, respectively.  (Red-solid) Flat
    spectrum expected in simple RD universe.  (Green-dashed) GW
    spectrum for the case without dark radiation.  Right: Blow-up of
    the left.  }
  \label{fig_411_twin}
\end{figure}
\begin{figure}
  \centerline{\epsfxsize=\textwidth\epsfbox{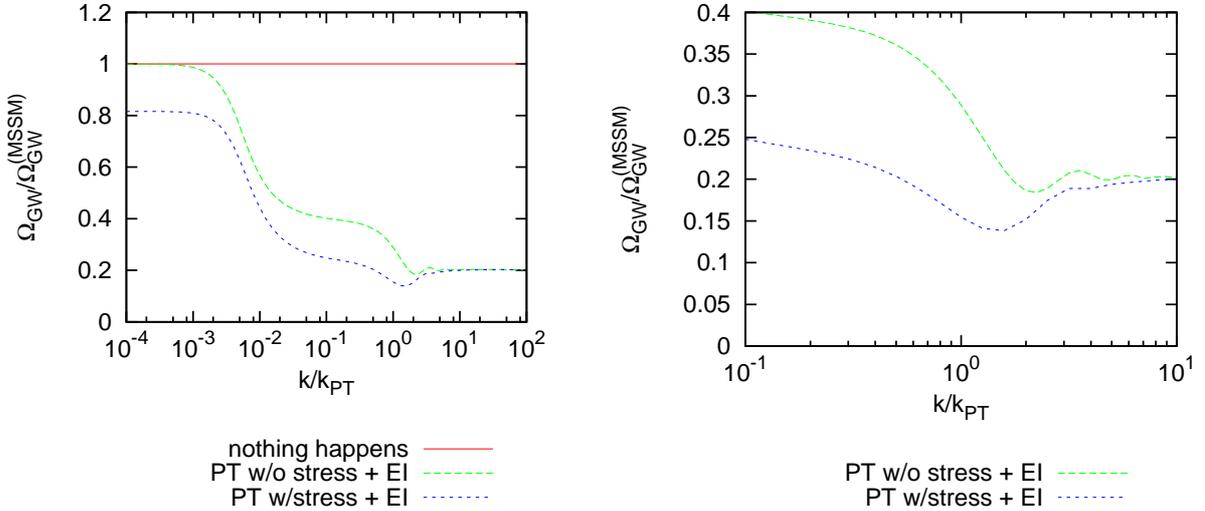}}
  \caption{\small Left: GW spectrum for Case 4. $R_{\rm
      coh}=0.01$ and $\Delta N_{\rm eff}=2$ just after the phase
    transition is assumed.  Right: Blow-up of the left.}
  \label{fig_412_twin}
\end{figure}

\subsection{Case 5}   \label{sec:case5}

In the final example, the scalar field $\phi$ remains as a coherent
oscillation after the phase transition.  Thus we assume the following
thermal history:
\begin{enumerate}
\item A brief period of thermal inflation is caused by a scalar field
  $\phi$.
\item After the phase transition, $\phi$ begins a coherent
  oscillation, which behaves as non-relativistic matter.
\item The coherent oscillation $\phi$ decays into dark and visible
  radiation.
\end{enumerate}
Then, the relevant evolution equations are:
\begin{eqnarray}
  &&\rho_{\rm vac} = \begin{cases}
    \Lambda^4 & ~~{\rm for}~~ t < t_{\rm PT}\\
    0                   & ~~{\rm for}~~ t > t_{\rm PT},
  \end{cases}
  \\
  &&\dot{\rho}_{\rm m} + 3H\rho_{\rm m} = -\Gamma\rho_{\rm m} + \Lambda^4 \delta(t-t_{\rm PT}) ,\\
  &&\dot{\rho}_{\rm r} + 4H \rho_{\rm r} = \Gamma B_{\rm r} \rho_{\rm m}, \\
  &&\dot{\rho}_{X} + 4H \rho_{X} = \Gamma B_{X} \rho_{\rm m},
\end{eqnarray}
where $B_{\rm r}$ and $B_X$ are branching ratio of $\phi$ into the
radiation and dark radiation, respectively.  They satisfy $B_{\rm r} +
B_{X} = 1$.

In \figref[fig_5_long] and \figref[fig_5_short], we show the resulting
GW spectrum.  In \figref[fig_5_long] we assumed a rather long period
of thermal inflation and coherent oscillation domination.  In the
figure, the parameters are
\begin{eqnarray}
R_{\rm coh} &=& 1, \nonumber \\
R_{\rm rad} &=& \left. \frac{\rho_{\rm r}}{\rho_{\rm tot}} \right|_{\rm just \; before \; PT} = 10^{-6}, \nonumber \\
R_{\rm dil} &=& \frac{\left. a^4 \rho_{\rm tot} \right|_{\rm just \; after \; PT}}{\left. a^4 \rho_{\rm tot} \right|_{\rm well \; after \; decay}}
= 10^{-6}, \nonumber
\end{eqnarray}
and $B_X = 0$. 

We can see that the GW spectrum scales as $k^{-2}$ if the horizon
entry is during the coherent-oscillation dominated era.  For the modes
which experience the horizon entry twice due to the thermal inflation,
the GW spectrum shows oscillatory behavior with its amplitude
proportional to $k^{-4}$.  (See Appendix \ref{sec:GW_inf}.)  We define
$k_{\rm d}$ as the transition frequency between these two regimes.
Note that $R_{\rm rad}$ parameterizes how much the
initial radiation has been diluted by the short inflation, and
therefore determines the ratio of the GW spectrum at $k \gg k_{\rm
  PT}$ to that at $k=k_{\rm d}$. 
On the other hand, $R_{\rm dil}$
gives how much the radiation which exists at the phase transition has
been diluted by the subsequent matter domination, and determines the
ratio of the GW spectrum at $k = k_{\rm d}$ to at $k \ll k_{\rm
  decay} \equiv aH(t=\Gamma^{-1})$.

\figref[fig_5_short], on the other hand, shows the GW spectrum with a
short period of thermal inflation and coherent oscillation domination.
Three lines (red-solid, green-dashed and blue-dotted) correspond to
the parameters
\begin{eqnarray}
R_{\rm coh} &=& 1, \nonumber \\
R_{\rm rad} &=& 0.67, \; 0.5, \; 0.33, \nonumber \\
R_{\rm dil} &=& 0.5, \nonumber
\end{eqnarray}
and we have taken $B_X$ so that we obtain $\Delta N_{\rm eff}=0$
$(0.5)$ at present in the left (right) figure. 
Although the spectral
shapes for the cases with and without dark radiation look similar,
the detailed structures are different.  Thus, 
precise observations of the height of the spectrum at $k \ll k_{\rm
  PT}$ and the amplitude of the oscillation around $k \simeq {\rm (a
  \; few)} \times k_{\rm PT}$ may make it possible to tell the
existence of the anisotropic stress.

\begin{figure}  
  \centerline{\epsfxsize=9cm \epsfbox{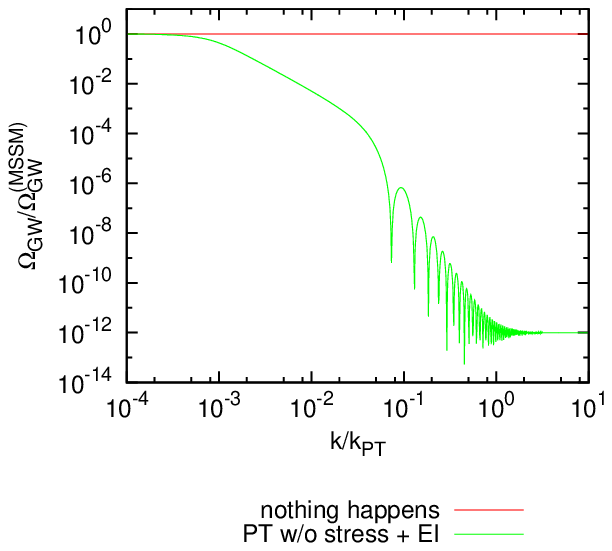}}
  \caption{\small GW spectrum for Case 5, taking $R_{\rm coh}=1,
    R_{\rm rad}=10^{-6}$ and $R_{\rm dil}=10^{-6}$.  }
  \label{fig_5_long}
  \centerline{\epsfxsize=\textwidth\epsfbox{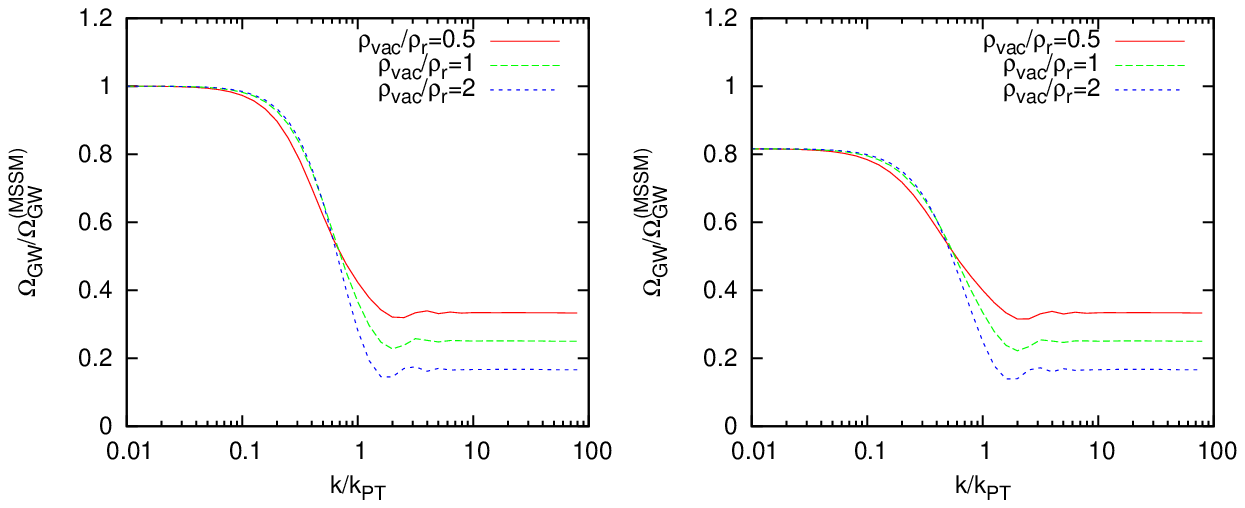}}
  \caption{\small Left: GW spectrum for Case 5 for $R_{\rm coh}=1,
    R_{\rm dil}=0.5$ and $R_{\rm rad}=0.67, 0.5, 0.33$ for the
    red-solid, green-dashed, blue dotted lines, respectively. We also
    assumed $\Delta N_{\rm eff} = 0$.  Right: Same as left but for
    $\Delta N_{\rm eff} = 0.5$.  }
  \label{fig_5_short}
\end{figure}

\section{Model}
\label{sec_Model}
\setcounter{equation}{0}

In this section we show that some particle-physics-motivated models
can realize the Cases 1--5 discussed in the previous section.  We take up two models for example, 
a SUSY PQ model and a SUSY majoron model.


\subsection{SUSY Peccei-Quinn model}

We consider a SUSY PQ model~\cite{Kawasaki:2013ae}.  In particular, we
consider the model with the following superpotential:
\begin{eqnarray}
W=\lambda S(\Phi\bar{\Phi}-f^2) + y_1\Phi Q_i\bar{Q}_i + y_2\bar{\Phi}Q'_j\bar{Q'}_j,
\label{eq_sp_SUSYaxino}
\end{eqnarray}
where $Q,Q'$ and $\bar{Q},\bar{Q'}$ are PQ quarks which are in the
fundamental and anti-fundamental representations of color SU(3),
respectively, and $f$ is the PQ scale.  The superpotential has a PQ
symmetry: $\Phi \rightarrow \Phi \e[i\alpha]$, $\bar{\Phi} \rightarrow
\bar{\Phi} \e[-i\alpha]$, $Q \rightarrow Q \e[-i\alpha]$ and $Q'
\rightarrow Q' \e[i\alpha]$.  The indices $i$ and $j$ run from 1 to
$a_1$ and $a_2$, respectively.\footnote
{ In order to avoid the axionic domain wall problem, we need
  $|a_2-a_1| = 1$ in the hadronic axion model~\cite{Hiramatsu:2010yn}.
}
In order for the PQ symmetry to be
anomalous under the color SU(3) for solving the strong CP problem, we
need $a_1 \neq a_2$.  For simplicity, we assume that the Yukawa
couplings $y_1$ and $y_2$ are independent of the indices $i$ and $j$.
The soft SUSY breaking potential is given by
\begin{eqnarray}
V_{\rm soft}
= m^2 \abs[\Phi]^2 + m^2 \abs[\bar{\Phi}]^2 + m^2 \abs[S]^2.
\end{eqnarray}
Here we have assumed that the soft masses $m$ are the
same for $\Phi, \bar \Phi$ and $S$ for simplicity and neglected
trilinear terms.\footnote
{Giving different soft masses to each field may shift the potential
  minimum and change axion decay constant by some factor, but the
  following qualitative argument does not change.  }
We rewrite the two superfields $\Phi$ and $\bar{\Phi}$ in terms of the
axion multiplet $A$ and heavy multiplet $A_H$ as
\begin{eqnarray}
&&\Phi = \braket{\Phi} + A \cos \theta + A_H \sin \theta, \\
&&\bar{\Phi} = \braket{\bar{\Phi}} - A \sin \theta + A_H \cos \theta,
\end{eqnarray}
where $\tan\theta = \braket{\bar{\Phi}}/{\braket{\Phi}}$.  The axion
$a$ and saxion $\sigma$ are embedded in the scalar component of $A$ as
$A = (\sigma + ia)/ \sqrt{2}$.  As long as the SUSY breaking mass is
much smaller than the PQ scale, $\Phi$ and $\bar{\Phi}$ are
constrained on $\Phi \bar{\Phi} = f^2$.

For sufficiently high cosmic temperature, the thermal effects keep
$\Phi$ and $\bar{\Phi}$ at the origin.  As the temperature decreases,
$\Phi$ and $\bar \Phi$ begin to roll down toward the minimum.  The
mass matrix for $\Phi$ and $\bar \Phi$ around the origin, including
thermal effects, is given by
\begin{eqnarray}
M^2 \sim \left(
\begin{matrix}
y^2 T^2 & -\lambda^2 f^2 \\
-\lambda^2 f^2 & y^2 T^2
\end{matrix}
\right),
\label{eq_TPT}
\end{eqnarray}
where $y\sim \sqrt{a_1}y_1 \sim\sqrt{a_2}y_2$.  Note that, if $a_1$
and $a_2$ are larger than 1, $y\gtrsim 1$ is possible even if
$y_1, y_2\lesssim 1$.  Thus the temperature at the time of phase transition
is evaluated as
\begin{eqnarray}
  T_{\rm PT} \sim \frac{\lambda}{y} f,
\end{eqnarray}
where, in our analysis, we assume that $y \gtrsim \lambda$. The ratio
of radiation energy density to the total one just before the phase
transition is then given by
\begin{eqnarray}
  R_{\rm rad} 
  =\frac{\frac{\pi^2}{30}g_{\ast}T_{\rm PT}^4}{\frac{\pi^2}{30}g_{\ast}T_{\rm PT}^4 + \lambda^2 f^4}.
\label{eq_Rrad}
\end{eqnarray}
If there is no production of dark radiation, the ratio $R_{\rm rad}$
also corresponds to the relative magnitude of the GW spectrum,
$\Omega_{\rm GW}(k \gg k_{\rm PT})/ \Omega_{\rm GW}(k \ll k_{\rm PT})$,
as discussed in the previous sections.

After the phase transition, we only consider the effects of the
particles in the axion multiplet $A$.  Other particles in the PQ
sector have masses larger than $T_{\rm PT}$ if $y \gtrsim 1$, which we
will assume in this subsection.  Then, we may eliminate $\bar{\Phi}$ in
terms of $\Phi$ by using $\Phi \bar{\Phi} = f^2$, and obtain the
potential for the scalar component of $\Phi$.  For the study of the
behavior of the axion multiplet in the early universe, it is important
to take account of various thermal
effects~\cite{Kawasaki:2010gv,Moroi:2012vu,Moroi:2013tea}.  In
particular, the effective potential of $\Phi$ is given in the
following form:
\begin{eqnarray}
  V=V_0+V_L+V_T.
\end{eqnarray}
Here, $V_0$ is the soft-mass term
\begin{eqnarray}
  V_0
  = m^2 \abs[\Phi]^2 + m^2 \abs[\bar{\Phi}]^2 
  = m^2 \abs[\Phi]^2 + m^2 \frac{f^4}{\abs[\Phi]^2}.
\end{eqnarray}
In addition, $V_L$ and $V_T$ represent thermal effects. $V_L$ is the
thermal log potential~\cite{Anisimov:2000wx}, which comes from the
fact that the masses of PQ quarks depend on the amplitude of $\Phi$:
\begin{eqnarray}
  V_L 
  \simeq 
  - \sum_{{\rm heavy \; quarks \;} Q} \alpha_L \alpha_3^2 T^4 \ln \abs[m_Q^2(\Phi)] 
  \simeq -2(a_2-a_1) \alpha_3^2 T^4 \ln \abs[\Phi].
\end{eqnarray} 
Note that $V_L$ is effective when the temperature is lower than the
masses of PQ quarks.  Furthermore, $V_T$ is the thermal potential
\begin{eqnarray}
  V_T = 
  \left\{
    \begin{array}{ll}
      \displaystyle{\frac{1}{24}} T^2m_{\tilde{Q}} (\Phi)^2 & 
      ~~~{\rm(for \; 1\; real \; scalar)}
      \\ \\
      \displaystyle{\frac{1}{24}} T^2 m_Q(\Phi)^2 &
      ~~~{\rm(for \; 1\; Weyl \; fermion)}
    \end{array}
  \right. ,
\end{eqnarray}
where $m_{\tilde{Q}}(\Phi)$ and $m_Q(\Phi)$ are masses of PQ squarks
and PQ quarks, respectively.  Note that $V_T$ is effective when the PQ
quarks are in the thermal bath, in contrast to $V_L$.  The minimum of
the potential depends on the temperature.  First, in the
low-temperature limit, $|\Phi|_{\rm min} \sim f$.  If the thermal log
potential is efficient, the minimum is given by $|\Phi|_{\rm min}\sim
\alpha_3T^2/m$.  Finally, in the high-temperature limit, the thermal
mass term determines the minimum at $|\Phi|_{\rm min} \sim (yT
f^2/m)^{1/2}$.

The trajectory of the PQ fields (in particular, $\Phi$ and
$\bar{\Phi}$) during the phase transition depends on $a_i$ and $y_i$.
Therefore it may be the case that a considerable fraction of the
initial vacuum energy may be transferred to the energy density of the
saxion coherent oscillation.  By using the parameter $R_{\rm coh}$
given in \eqref[R_coh], we phenomenologically parameterize the
fraction of $\rho_{\rm vac}$ transferred into the coherent oscillation
energy density $\rho_{\rm coh}$; the fraction $1 - R_{\rm coh}$ of
$\rho_{\rm vac}$, which corresponds to the energy density of the
oscillation of the field transverse to the trajectory
$\Phi\bar{\Phi}=f^2$, is assumed to be instantly transferred into the
energy densities of the axion and saxion:
\begin{eqnarray}
\rho_{\rm a} \sim \rho_{\rm \sigma} \sim \frac{1-R_{\rm coh}}{2} \rho_{\rm vac}.
\end{eqnarray}
Hereafter we consider the case where the initial vacuum energy does not go into the visible radiation.
If $R_{\rm coh}$ is not negligible, the amplitudes of $\Phi$ or
$\bar{\Phi}$ can take large value due to the dynamics along the saxion
direction.  The maximal value of $\abs[\Phi]$ during the coherent
oscillation is
\begin{eqnarray}
  \abs[\Phi]_{\rm max}
  \simeq \frac{\lambda f^2 R_{\rm coh}^{1/2}}{m}.
  \label{eq_coh_max}
\end{eqnarray}
Using $m_{Q'} = y f^2/ \abs[\Phi]$ and $T_{\rm PT}\sim\lambda f/y$, the
PQ quarks are not in thermal equilibrium if
\begin{eqnarray}
\left( \frac{y}{\lambda} \right)^2 > \frac{f}{m}.
\label{eq_qmassive}
\end{eqnarray}
If this condition is satisfied, the saxion may be trapped at the local
minimum and the onset of the saxion coherent oscillation may be
delayed.  

The subsequent evolution of the system depends on the decay rate and
dissipation rate of the fields.  
First, let us consider the collision rate of the particles in the PQ
sector.  Without PQ quarks in the thermal bath, the axion, saxion, and
axino hardly communicate with the visible sector particles.  Then,
they may be sequestered from the visible sector and the temperature of
the PQ-sector particles may be different from that of the
visible-sector particles.  In particular, if the temperature of the PQ
sector particles, denoted as $T^{\rm (PQ)}$, is lower than $\sim
\lambda f$, the collision rate among the PQ sector particles is
approximately given by
\begin{eqnarray}
\Gamma_{\rm coll}\simeq 4 \times 10^{-4}\ \frac{T^{{\rm (PQ)}5}}{f^4}.
\end{eqnarray}
Thus, when $T^{\rm (PQ)}$ becomes lower than the decoupling
temperature $T_{\rm decouple}^{\rm (PQ)}$, the PQ-sector particles
behave as dark radiation (as long as they are relativistic), where
$T_{\rm decouple}^{\rm (PQ)}$ is estimated as
\begin{eqnarray}
  T_{\rm decouple}^{\rm (PQ)}
  \sim 
  2\times 10^{11}\ {\rm GeV}
  \times
  \left( \frac{f}{10^{12}\ {\rm GeV}} \right)^{4/3}.
\end{eqnarray}
Here the ratio of the energy density of the visible sector to the
invisible one is $R_{\rm rad}/(1-R_{\rm rad})$ in the absence of the
saxion coherent oscillation, therefore the temperature of the visible
sector at the time of the decoupling of the PQ-sector particles is
\begin{eqnarray}
  T_{\rm decouple}
  = 0.36 \times \left( \frac{R_{\rm rad}}{1-R_{\rm rad}} \right)^{1/4} 
  T_{\rm decouple}^{\rm (PQ)}.
\end{eqnarray}

We also comment here that the effect of dissipation may be important
in studying the evolution of the saxion in thermal bath
\cite{Moroi:2012vu}.  When PQ quarks are not in thermal equilibrium
($m_Q \gtrsim T$), the dissipation rate of the axion, saxion and axino
via the scattering with the gluon is \cite{Laine:2010cq}
\begin{eqnarray}
  \Gamma_{\rm diss}
  \sim \frac{9\alpha_3^2}{128\pi^2 \ln \alpha_3^{-1}}
  \frac{T^3}{{\rm max}(\Phi,\bar{\Phi})^2}.
\end{eqnarray}
When PQ quarks are in thermal bath, on the other hand, (s)axions
collide with particles in the heavier multiplets with the rate
\begin{eqnarray}
\Gamma_{\rm diss} \sim 10^{-2} \times y^4 T.
\end{eqnarray}
With the choices of parameters for the following discussion, however,
$\Gamma_{\rm diss}$ is smaller than the expansion rate of the universe
so that the effects of dissipation are irrelevant.  It means that the
scattering rate between the PQ sector and visible sector is so small
that the PQ sector is sequestered from the visible sector.  Thus we
can define the temperature of the PQ sector and the visible sector
separately.

After the decoupling, the saxion decays at the late epoch.  The
partial decay rates of the saxion into axion and gluon pairs are
\begin{eqnarray}
\Gamma_{\sigma \rightarrow aa}
&\simeq& \frac{1}{64\pi}\frac{m^3}{f^2},
\label{Gamma_sigma2aa}\\
\Gamma_{\sigma \rightarrow gg}
&\simeq& \frac{\alpha_3^2}{32\pi^3}\frac{m^3}{f^2},
\end{eqnarray}
respectively, where, in deriving \eqref[Gamma_sigma2aa], we assumed that the
difference of the vacuum expectation values of $\Phi$ and $\bar{\Phi}$
is sizable.  In addition, if $\Phi$ couples to the Higgs doublets, as
$\Phi H_uH_d$, for example, and also if the higgsino mass is
dominantly from the vacuum expectation value of $\Phi$, the saxion may
also decay into the Higgs-boson pair with the following decay rate:
\begin{eqnarray}
\Gamma_{\sigma \rightarrow HH}\simeq 
\frac{1}{2\pi}\frac{m^3}{f^2}
\left( \frac{\mu}{m} \right)^4,
\end{eqnarray}
where $\mu$ is the higgsino mass.  
The decay temperature of the saxion is estimated by solving the
following equation: 
\begin{eqnarray}
  \frac{m}{\langle E_\sigma \rangle(T_{\rm decay})}
  \Gamma_{\sigma}
  \sim 3H(T_{\rm decay}),
\end{eqnarray}
where $\langle E_\sigma\rangle(T)$ is the averaged energy of saxion at
the cosmic temperature $T$.  The decay temperature depends on the
dominant decay mode of the saxion, and also on the typical energy of the saxion at the time of the decay. 
If $R_{\rm coh} \sim 0$ and $R_{\rm rad} \sim 0$ and the saxion dominantly decays into the Higgs pair, 
the temperature of the visible sector $T_{\rm decay}$ just after the decay is estimated as
\begin{eqnarray}
  T_{\rm decay} \sim 4 \times 10^3\ {\rm GeV} \times 
  \left( \frac{\mu}{10^5\,{\rm GeV}} \right)^{4/3}
  \left( \frac{f}{10^{12}\,{\rm GeV}} \right)^{-2/3}.
\end{eqnarray}
where we approximated that $\langle E_\sigma \rangle\sim T^{\rm (PQ)}$
and used $g_{*}^{\rm (PQ)}=3.75$ as the effective number of relativistic
degrees of freedom in the PQ sector, and also assumed that the saxion decays while being relativistic.
If the saxion decays after becoming non-relativistic, we obtain
\begin{eqnarray}
  T_{\rm decay} 
  \sim 5 \times 10^3\ {\rm GeV} \times 
  \left( \frac{\mu}{10^5\,{\rm GeV}} \right)^{2}
  \left( \frac{m}{10^5\,{\rm GeV}} \right)^{-1/2}
  \left( \frac{f}{10^{12}\,{\rm GeV}} \right)^{-1}.
\end{eqnarray}

Now let us see that the SUSY axion model can realize the Cases 1 and 2
studied in \secref[sec:case1] and \secref[sec:case2], respectively.
The role of $\phi$ is played by the PQ scalar, $\Phi$ and $\bar\Phi$.
After the PQ phase transition, they decay into particles in the axion
multiplet, which are thermalized due to their self interactions.
(Here, we assume that the thermal bath of the PQ sector particles are
sequestered from that of the visible sector.)  At some point, the
interactions of the PQ sector particles become very weak.  Then, the
mean free path of the axion and saxion becomes so long that they
behave as dark radiation $X$.  After some periods, the saxion decays
into radiation.  If the saxion decays while it is relativistic, it
corresponds to the Case 1 and if it is non-relativistic, the Case 2 is
realized.

A sample parameter set corresponding to the Case 1 is
\begin{eqnarray}
f&=&10^{12}\,{\rm GeV}, \nonumber \\
\mu &=& m = 10^6\,{\rm GeV}, \nonumber \\
\lambda&=& 1, \nonumber \\
y&=& 2, \nonumber \\
R_{\rm coh}&=&0.
\end{eqnarray}
With the above parameters, we obtain
\begin{eqnarray}
T_{\rm PT} &\simeq& 10^{12}\, {\rm GeV}, \nonumber \\
T_{\rm decouple} &\simeq& 10^{11}\, {\rm GeV}, \nonumber \\
T_{\rm decay} &\simeq& 10^6\, {\rm GeV}, \nonumber \\
R_{\rm rad}&\simeq&0.8.
\end{eqnarray}
The cosmological evolution goes as follows.
\begin{itemize}
\item After the phase transition, the radial oscillation of $\Phi$ and
  $\bar\Phi$ quickly decays into the particles in the axion
  multiplet.
\item After the temperature drops down to $T \simeq T_{\rm decouple}$,
  the collision rate in the PQ sector becomes smaller than the Hubble
  expansion rate.  So axion, saxion and axino freely stream and hence
  behave as dark radiation.
\item If the field $\Phi$ or $\bar{\Phi}$ couples to the
  Higgs sector, the saxion can mainly decays into the Higgs bosons.
  The saxion decays at $T = T_{\rm dec} \simeq 10^6$\,GeV before it
  becomes massive.  This eliminates about one-third of dark
  radiation. If the axino also decays without dominating the universe,
  $\Delta N_{\rm eff} \sim 1$ at the present universe.
\end{itemize}
The features of the GW spectrum expected with the above setup are the
same as \figref[fig_1_2_2], but the positions of the characteristic
wavenumber are different.  In \figref[fig_1_2_2], $T_{\rm decouple} =
10^{-2}T_{\rm PT}$ and $T_{\rm decay} = 10^{-4}T_{\rm PT}$ is assumed
while in the present setup they become $T_{\rm decouple} = 10^{-1}
T_{\rm PT}$ and $T_{\rm decay} = 10^{-6} T_{\rm PT}$.

A sample parameter set corresponding to the Case 2 is,
\begin{eqnarray}
f&=&10^{12}\, {\rm GeV}, \nonumber \\
\mu &=& m = 10^5\, {\rm GeV}, \nonumber \\
\lambda&=&1, \nonumber \\
y&=& 2, \nonumber \\
R_{\rm coh}&=&0,
\end{eqnarray}
which gives
\begin{eqnarray}
T_{\rm PT} &\simeq& 10^{12}\, {\rm GeV}, \nonumber \\
T_{\rm decouple} &\simeq& 10^{11}\, {\rm GeV}, \nonumber \\
T_{\rm decay} &\simeq& 10^4\, {\rm GeV}, \nonumber \\
R_{\rm rad}&=&0.8.
\end{eqnarray}
The cosmological evolution goes as follows.
\begin{itemize}
\item After the phase transition, the radial oscillation of $\Phi$ and
  $\bar\Phi$ quickly decays into the PQ sector.
\item After the temperature drops down to $T \simeq T_{\rm decouple}$,
  the collision rate in the PQ sector becomes smaller than the Hubble
  expansion rate, and the axion, saxion and axino become dark radiation.
\item The saxion becomes non-relativistic at $T \simeq m =10^5$\,GeV.
  Then, it decays into the Higgs bosons at $T_{\rm dec} \simeq
  10^4$\,GeV.
\end{itemize}
The features in the GW spectrum with this setup are expected to be the
same as in \figref[fig_2_1_3_2] and \figref[fig_2_2_3_2], although the
characteristic wavenumbers are different.

If a large amount of the vacuum energy once goes into the coherent
oscillation of the saxion, the Case 5 may also be realized in the SUSY
PQ model.  With a proper choice of parameters, most of the vacuum
energy goes to the coherent oscillation of the saxion.  The saxion
dominates the universe and then decays; significant amount of the
axion, which plays the role of dark radiation, may be produced by the
decay with a relevant choice of parameters.

In this setup, the initial radiation is first diluted by the vacuum
energy just like in the previous cases, and then diluted by the matter
(i.e., saxion) domination which begins just after the phase
transition.  The former dilution can be parameterized by $R_{\rm
  rad}$ previously defined.  Let us define the quantity $R_{\rm dil}$,
which parameterize the latter dilution:
\begin{eqnarray}
  R_{\rm dil} 
  &\equiv& 
  \frac{\left. a^4 \rho_{\rm tot} \right|_{\rm just \; after \; PT}}
  {\left. a^4 \rho_{\rm tot} \right|_{\rm well \; after \; decay}} 
  \nonumber \\ &\simeq& 
  2 \times 10^{-11}\ \lambda^{-4/3} y^{4/3} 
  \left( \frac{\mu}{10^5\ {\rm GeV}} \right)^{8/3}
  \left( \frac{m}{10^5\ {\rm GeV}} \right)^{-2/3} 
  \left( \frac{f}{10^{12}\ {\rm GeV}} \right)^{-8/3}.
\end{eqnarray}
This determines the ratio of the spectral height at $k=k_{\rm d}$ to
that at $k \ll k_{\rm decay} \equiv aH(t=\Gamma^{-1})$.

The first example in the Case 5 discussed in \secref[sec:case5] is
realized with the following parameters:
\begin{eqnarray}
f&=& {\rm (a \; few)} \times 10^{11}\, {\rm GeV}, \nonumber \\
\mu &=& m = 10^6\, {\rm GeV}, \nonumber \\
\lambda &=&10^{-3}, \nonumber \\
y &=& 1, \nonumber \\
R_{\rm coh} &=& 1,
\end{eqnarray}
which result in
\begin{eqnarray}
T_{\rm PT} &\simeq& 10^8\, {\rm GeV}, \nonumber \\
R_{\rm rad} &\simeq& 10^{-4}, \nonumber \\
R_{\rm dil} &\simeq& 10^{-4}.
\end{eqnarray}
Thus, the cosmological evolution goes as follows:
\begin{itemize}
\item The thermal inflation caused by the vacuum energy of $\Phi$ and
  $\bar{\Phi}$ dilutes the initial visible radiation by a factor of
  $10^{-4}$.
\item After the phase transition, almost all the vacuum energy goes to
  the saxion coherent oscillation.  The heavy quarks remain massive
  since \eqref[eq_qmassive] is satisfied.
\item After the coherent oscillation dilutes the radiation further by
  a factor of $10^{-4}$, it decays into visible radiation.
\end{itemize}

The second example in the Case 5 is realized with the following
parameters:
\begin{eqnarray}
f&=& {\rm (a \; few)} \times 10^{9}\, {\rm GeV}, \nonumber \\
\mu &=& m = 10^6\, {\rm GeV}, \nonumber \\
\lambda &=&10^{-1}, \nonumber \\
y &=& 1, \nonumber \\
R_{\rm coh} &=& 1,
\end{eqnarray}
which give
\begin{eqnarray}
T_{\rm PT} &\simeq& 10^8\, {\rm GeV}, \nonumber \\
R_{\rm rad} &=& {\mathcal O} (0.1), \nonumber \\
R_{\rm dil} &=& {\mathcal O} (0.1). \nonumber 
\end{eqnarray}
The cosmological evolution goes as follows:
\begin{itemize}
\item The thermal inflation caused by the vacuum energy of $\Phi$ and
  $\bar{\Phi}$ dilutes the initial visible radiation by a factor of
  ${\mathcal O}(0.1)$.
\item After the phase transition, almost all the vacuum energy goes to
  the saxion coherent oscillation.  The heavy quarks remain massive
  since \eqref[eq_qmassive] is satisfied.
\item The coherent oscillation dilutes the radiation by a factor of
  ${\mathcal O}(0.1)$, and it soon decays into visible radiation.
\end{itemize}

\subsection{SUSY majoron model}

Next, we consider a SUSY majoron
model~\cite{Chikashige:1980ui,Chun:1994zp}.  The mass of the
right-handed neutrino, which is needed for the seesaw mechanism
\cite{seesaw}, can be generated as a consequence of U(1) symmetry
breaking.  This U(1) may be identified as a gauged $B-L$ symmetry, or
it can be global lepton number symmetry. Here we consider the latter
option.  Then a Nambu-Goldstone (NG) boson appears which is associated
with the spontaneous breakdown of the U(1) global symmetry.  This NG
boson is called majoron.  (We call other fields in the majoron
supermultiplet as smajoron, which is a real scalar field, and
majorino, which is the fermionic superpartner of the majoron.)

Let us consider the following superpotential as an example of a SUSY
majoron model,
\begin{eqnarray}
  W=\lambda S(\Phi\bar{\Phi}-f^2) + y_i \Phi N_i N_i 
  + y_i' \bar\Phi N_i' N_i',
\end{eqnarray}
where $N_i$ are the right-hand neutrinos, while $\Phi$ and
$\bar{\Phi}$ are lepton-number violating fields.  In the following
argument, we take $y_i = y'_i = y$ for simplicity.  In addition, we
also introduce $N_i'$ which have opposite lepton number.  In contrast
to the SUSY PQ model, this model does not lead to the thermalization
of the majoron sector with the SM sector since the neutrino Yukawa
coupling constants are taken to be small enough.  Note also that the
parameter $f$, the U(1) symmetry breaking scale, is not severely
constrained in contrast to the PQ scale.

In the following, we assume that the majoron and the right-handed
neutrinos are initially in the thermal bath.  (We use the same
notation as in the previous section for $T_{\rm PT}$, $T_{\rm
  decouple}$, $T_{\rm decay}$, $R_{\rm coh}$ and $R_{\rm
  rad}$.)

Now let us see that the SUSY majoron model can realize the Case 3 and
4 studied in \secref[sec:case3] and \secref[sec:case4].  For the Case
3, we take the following parameters:
\begin{eqnarray}
f&=&10^{10}\,{\rm GeV}, \nonumber \\
\mu &=& m = {\rm (a \;few)} \times 10^5\,{\rm GeV}, \nonumber \\
\lambda&=& 10^{-3}, \nonumber \\
y&=& 10^{-1}, \nonumber \\
R_{\rm coh}&=&0.
\end{eqnarray}
Then, we obtain
\begin{eqnarray}
T_{\rm PT} &\simeq& 10^{8}\, {\rm GeV}, \nonumber \\
T_{\rm decay} &\simeq& 10^6\, {\rm GeV}.
\end{eqnarray}
The cosmological evolution goes as follows:
\begin{itemize}
\item After the phase transition, the radial oscillation of $\Phi$ and
  $\bar\Phi$ quickly decays into the majoron sector.  The majoron
  sector does not thermalize because of the weakness of the
  interaction of the majoron.
\item The smajoron decays at $T = T_{\rm dec} \simeq 10^6$\,GeV before
  it becomes massive.  This eliminates about one-half of the dark
  radiation, realizing the present $\Delta N_{\rm eff} = {\mathcal
    O}(0.1)$.
\end{itemize}

For the Case 4, we take the following parameters:
\begin{eqnarray}
f&=&10^{11}\,{\rm GeV}, \nonumber \\
\mu &=& m = 10^6\,{\rm GeV}, \nonumber \\
\lambda&=& 10^{-5}, \nonumber \\
y&=& 10^{-2}, \nonumber \\
R_{\rm coh}&=&0.01,\; 0.1.
\end{eqnarray}
The value of $R_{\rm coh}$ corresponds to \figref[fig_411_twin] and
\figref[fig_412_twin], respectively.  With the above choice of
parameters,
\begin{eqnarray}
T_{\rm PT} &\simeq& 10^8\, {\rm GeV}, \nonumber \\
T_{\rm decay} &\simeq& 10^6\, {\rm GeV}.
\end{eqnarray}
The cosmological evolution goes as follows:
\begin{itemize}
\item After the phase transition, a small fraction of the vacuum
  energy goes into the coherent oscillation of the smajoron.  The
  rest, the radial oscillation of $\Phi$ and $\bar\Phi$, quickly
  decays into the majoron sector particles.  The majoron sector does
  not thermalize.
\item The coherent oscillation of the smajoron begins to dominate the
  universe after the temperature decreases to $10^{-1} T_{\rm PT}$ and
  $10^{-2} T_{\rm PT}$ in \figref[fig_411_twin] and
  \figref[fig_412_twin], respectively.  This dilutes the dark
  radiation produced at the phase transition.  Note that the smajoron
  particles are still relativistic or just becoming non-relativistic at these
  temperatures.
\item The coherent oscillation (and smajoron particles in the case of
  \figref[fig_412_twin],) decays into visible radiation and the
  present $\Delta N_{\rm eff}$ becomes ${\mathcal O}(0.1)$.
\end{itemize}

\section{Conclusions}
\label{sec_conc}

In this paper we have studied various effects of the cosmological
events in the early universe on the inflationary GW spectrum.  In
particular, some (well-motivated) models of cosmological phase
transition, including the PQ phase transition, in general lead to
complicated structures in the GW spectrum, which might be detectable
in future space-based GW
detectors~\cite{Seto:2001qf,Crowder:2005nr,Cutler:2009qv}.  Hence, if
the spectrum of the inflationary GWs is determined by future
experiments, it will give clues to the high-energy physics beyond the
reach of accelerator experiments.

Finally we mention inflation models which predict observable levels of
GWs.  Although the recent Planck results do not favor the standard
chaotic inflation model with a simple power-law
potential~\cite{Ade:2013zuv}, a class of large field models has been
proposed which fits the Planck results well while predicts the
tensor-to-scalar ratio of $r\sim \mathcal
O(0.01)$~\cite{Croon:2013ana,Nakayama:2013jka} (see
Refs.~\cite{Linde:1984cd} for early attempts).  The Higgs
inflation~\cite{Bezrukov:2007ep,Ferrara:2010yw,Kallosh:2013pby}
predicts $r\sim \mathcal O(10^{-3})$, which is within the reach of
future GW detectors.\footnote{ For other types of Higgs inflation, see
  Refs.~\cite{Germani:2010gm,Nakayama:2010sk,Kamada:2012se}.  } The
$R^2$-inflation model~\cite{Starobinsky:1980te} (see
Refs.~\cite{Ellis:2013xoa,Kallosh:2013lkr} for its generalization) and
topological
inflation~\cite{Linde:1994hy,Vilenkin:1994pv,Harigaya:2012hn} also
predict observable GWs with $r\sim \mathcal O(10^{-3})$.

\section*{Acknowledgment}

We thank K.~Schmitz for comments regarding footnote~\ref{footnote:3}.
This work is supported by Grant-in-Aid for Scientific
research from the Ministry of Education, Science, Sports, and Culture
(MEXT), Japan, No.\ 22244021 (T.M.), No.\ 22540263 (T.M.), No.\
23104001 (T.M.), No.\ 21111006 (K.N.), and No.\ 22244030 (K.N.).
The work of R.J. is supported in part by JSPS Research Fellowships
for Young Scientists.

\appendix

\section{Inflationary GW power spectrum} 
\label{app:tensor}
\setcounter{equation}{0}

The dimensionless power spectrum $\Delta^2(k)$ of some perturbation $\delta(\mb[x])$ which obeys a homogeneous and 
isotropic distribution is defined as
\begin{eqnarray}
\braket{\delta(\mb[x])\delta(\mb[x]')}
= \int \frac{d^3k}{(2\pi)^3} e^{i \mb[k]\cdot\mb[x] } P_\delta(k)
= \int \frac{dk}{k} \Delta^2(k) \frac{\sin(kr)}{kr},
\end{eqnarray}
where $r=|\mb[x]-\mb[x]'|$ and the bracket means taking ensemble average, which 
we assume to be the same as spacial average. 
In terms of Fourier-transformed perturbation $\delta(\mb[k])$, $\Delta(k)$ is written as
\begin{eqnarray}
\Delta^2(k)
= \frac{k^3}{2\pi^2} P_\delta (k),
\end{eqnarray}
where
\begin{equation}
	\braket{\delta(\mb[k])\delta(\mb[k]')} \equiv (2\pi)^3 \delta(\mb[k]+\mb[k]') P_\delta(k).
\end{equation}
In the present case $\delta(\mb[x])$ corresponds to $h_{ij}(t,\mb[x])$:
\begin{eqnarray}
\Delta_{\rm GW}^2(t,k) 
= \frac{k^3}{2 \pi^2} P_h(t,k),
\end{eqnarray}
where
\begin{equation}
	\sum_{\lambda=+,\times}\braket{h(t,\mb[k],\lambda)h(t,\mb[k]',\lambda)} \equiv (2\pi)^3 \delta(\mb[k]+\mb[k]') P_h(t,k).
\end{equation}


In the inflationary era the perturbations with wavenumbers of our interest go far out of the horizon because of the 
exponential growth of the scale factor. Since then $h(t,\mb[k],\lambda)$ is constant until it re-enters the horizon, 
and hence $\Delta_{\rm GW}^2(t,k)$ is constant in time outside the horizon:
\begin{eqnarray}
\Delta_{\rm GW}^2(t,k)
= \Delta_{\rm GW,prim}^2(k) ~~~{\rm for}~~k \ll aH.
\end{eqnarray}
Here ``prim'' means its primordial value, evaluated at the time after
the horizon exit.  On the other hand, as shown in
Sec.~\ref{sec_Properties}, the amplitude of GWs decreases as $a^{-1}$
inside the horizon $(k \gg aH)$.  Hence,
\begin{eqnarray}
\left\langle \Delta_{\rm GW}^2(t,k) \right\rangle_{\rm osc}
= \frac{1}{2} \Delta_{\rm GW,prim}^2(k)\left(\frac{a_{\rm hi}(k)}{a(t)} \right)^2  ~~~{\rm for}~~k \gg aH,
\label{app:DeltaGW}
\end{eqnarray}
where $\braket{\cdots}_{\rm osc}$ is for oscillation average.  Here,
$a_{\rm hi}(k)$ is the scale factor at which the mode $k$ enters the
horizon $k/H = a_{\rm hi}(k)$.  Note the factor $1/2$ here is from the
fact that $h$ is oscillating with high frequency inside the horizon.\footnote
{ Here we have considered GWs entering the horizon at the RD era.  For
  those entering in the MD era, the factor $1/2$ in
  Eq.~(\ref{app:DeltaGW}) should be replaced with $9/32$, which is
  derived from the analytical solution given in Eq.~(\ref{h_MD}).  }

Assuming a slow-roll inflation and the canonical commutation relation
of $h_{ij}$, one obtains~\cite{Maggiore:1999vm}
\begin{eqnarray}
\Delta_{\rm GW,prim}^2(k)
= 64\pi G \left( \frac{H_{\rm ho}(k)}{2 \pi} \right)^2,
\end{eqnarray}
where $H_{\rm ho}$ is the value of $H$ at the time of horizon exit $k=aH$. 
Because GWs of different wavenumbers exit the horizon at different epochs, $H_{\rm ho}$ depends on $k$:
\begin{eqnarray}
H_{\rm ho}^2(k)
= H_{\rm ho}^2(k_0) \left( \frac{k}{k_0} \right)^{n_t},
\end{eqnarray}
where $k_0 = 0.002$Mpc${}^{-1}$ is the pivot scale and $n_t$ is the tensor spectral index.

Now, we relate the energy density of inflationary GWs to $\Delta_{\rm
  GW}^2$.  We define the effective energy-momentum tensor of GWs as
\begin{eqnarray}
  T_{\mu \nu}^{\rm GW} (t,\mb[x])
  = -\frac{1}{32\pi G} \braket{h^{ij}_{\;\;\; ; \mu} h_{ij;\nu}}_{\rm osc}
  (t,\mb[x]),
\end{eqnarray}
where $\braket{\cdots}_{\rm osc}$ here indicates the oscillation
average as well as the ensemble average; note that this expression is
applicable only to GWs with subhorizon modes $k \gg aH$. The effective
energy density of GWs is then defined as
\begin{eqnarray}
\rho_{\rm GW} (t,\mb[x])
\equiv {\left[ T_{\rm GW} \right]^0}_0 (t,\mb[x])
= \frac{1}{32\pi G} \braket{h^{ij;0} h_{ij;0}}_{\rm osc} (t,\mb[x]).
\label{eq_app_GW_energy}
\end{eqnarray}
We define the effective energy of GWs per logarithmic frequency
$\rho_{\rm GW}(t,k)$ through
\begin{eqnarray}
  \rho_{\rm GW}(t)
  \equiv \int d\ln k \; \rho_{\rm GW} (t,k),
\label{eq_app_GW_logenergy}
\end{eqnarray}
where we have assumed the homogeneity and isotropy of the primordial
GWs.  Because of the subhorizon condition $k \gg aH$, we may neglect
terms with time derivative on the scale factor in
\eqref[eq_app_GW_energy].  If the universe is filled with
perfect fluid at $t$, \eqref[eq_GW_eom_k] tells us that $h$ is a
harmonic oscillator with frequency $k/a$.  Thus for subhorizon modes,
we can use the relation $\langle\dot{h}_{ij}^2\rangle_{\rm osc}=
\langle(k/a)^2h_{ij}^2\rangle_{\rm osc}$ to write $\rho_{\rm GW}(t,k)$
as
\begin{eqnarray}
\rho_{\rm GW}(t,k)
= \frac{1}{32 \pi G} \frac{k^2}{a^2} \frac{k^3}{2 \pi^2} P_h(t,k)
= \frac{1}{32 \pi G} \frac{k^2}{a^2} \Delta_{\rm GW}^2(t,k).
\label{eq_app_GW_energy_delta}
\end{eqnarray} 

The GW spectrum $\Omega_{\rm GW}(t,k)$ is defined as the fraction of
$\rho_{\rm GW}(t,k)$ to the critical density:
\begin{eqnarray}
\Omega_{\rm GW}(t,k)
\equiv \frac{\rho_{\rm GW}(t,k)}{\rho_{{\rm tot}}(t)}.
\end{eqnarray}
The present value of $\Omega_{\rm GW}$ may be decomposed as follows
\begin{eqnarray}
\Omega_{\rm GW}(t_0,k)
=
\frac{\rho_{\rm GW}(t_0,k)}{\rho_{\rm GW}(t_{\rm hi},k)} 
\frac{\rho_{\rm GW}(t_{\rm hi},k)}{\rho_{\rm r}(t_{\rm hi})}
\frac{\rho_{\rm r}(t_{\rm hi})}{\rho_{\rm r}(t_0)}
\frac{\rho_{\rm r}(t_0)}{\rho_{\rm tot}(t_0)}.
\end{eqnarray}
Assuming that no significant entropy injection has occurred between
the time of horizon entry for GWs with wavenumber $k$ and the
matter-radiation equality, we obtain
\begin{eqnarray}
\frac{\rho_{\rm GW}(t_0,k)}{\rho_{\rm GW}(t_{\rm hi},k)} 
&\simeq& \left( \frac{a_{\rm hi}}{a_0} \right)^4, 
\\
\frac{\rho_{\rm GW}(t_{\rm hi},k)}{\rho_{\rm r}(t_{\rm hi})}
&=& \frac{1}{64\pi G} \frac{k^2}{a_{\rm hi}^2} \Delta_{\rm GW,prim}^2 (k) \frac{1}{\rho_{\rm r,hi}} = \frac{1}{24} \Delta_{\rm GW,prim}^2 (k_0) \left(\frac{k}{k_0}\right)^{n_t},\\
\frac{\rho_{\rm r}(t_{\rm hi})}{\rho_{\rm r}(t_0)}
&=&  \frac{\frac{\pi^2}{30}g_{\ast}(T_{\rm hi})T_{\rm hi}^4}{ \frac{\pi^2}{30} g_{\ast}(T_0) T_0^4}
= \left( \frac{g_{\ast}(T_{\rm hi})}{g_{\ast}(T_{\rm eq})} \right) \left( \frac{g_{\ast s}(T_{\rm eq}) a_0^3}{g_{\ast s}(T_{\rm hi}) a_{\rm hi}^3} \right)^{4/3}, \\
\frac{\rho_{\rm r}(t_0)}{\rho_{\rm tot}(t_0)}
&=& \Omega_{{\rm r},0},
\end{eqnarray}
where $\Omega_{{\rm r},0}$ and $g_*(T_0)$ are defined as if all neutrinos would be relativistic: $g_*(T_0)=g_*(T_{\rm eq})$.
The final numerical result does not depend on this definition. See also Ref.~\cite{Buchmuller:2013lra}.
With the tensor-to-scalar ratio $r$, which is defined as
\begin{eqnarray}
  r \equiv \frac{\Delta_{\rm GW,prim}^2(k_0)}{\Delta_{\rm{\cal R},prim}^2(k_0)},
\end{eqnarray}
where $\Delta_{\rm{\cal R},prim}^2$ is the dimensionless power spectrum of curvature perturbation ${\cal R}$, $\Omega_{\rm GW}$ becomes
\begin{eqnarray}
\Omega_{\rm GW}(t_0,k)
= \frac{r}{24} \Omega_{\rm r,0} \Delta_{\rm {\cal R},prim}^2(k_0) \left( \frac{k}{k_0} \right)^{n_t}
\left( \frac{g_{\ast}(T_{\rm hi})}{g_{\ast}(T_{\rm eq})} \right) \left( \frac{g_{\ast s}(T_{\rm eq})}{g_{\ast s}(T_{\rm hi})} \right)^{4/3}.
\end{eqnarray}
Substituting the observed values of the radiation fraction
$\Omega_{\rm r,0}\simeq 8.55 \times 10^{-5} [g_*(T_{\rm eq})/g_*(T_{\rm eq})^{\rm (std)}]$ and the magnitude of the primordial curvature 
perturbation $\Delta_{\rm {\cal R},prim}^2 (k_0)\simeq 2.22 \times 10^{-9}$~\cite{Ade:2013zuv}, one obtains the expression for the present GW spectrum:
\begin{eqnarray}
\Omega_{\rm GW}(t_0,k)
\simeq 7.9 \times 10^{-15} \left( \frac{g_*(T_{\rm hi})}{g_*(T_{\rm eq})^{\rm (std)}} \right) 
\left( \frac{g_{*s}(T_{\rm eq})}{g_{*s}(T_{\rm hi})} \right)^{4/3} \left( \frac{k}{k_0} \right)^{n_t} r.
\end{eqnarray}
In the MSSM, it becomes
\begin{eqnarray}
\Omega_{\rm GW}^{\rm (std)}(t_0,k)
\simeq 2.3 \times 10^{-15} \left( \frac{228.75}{g_{*s}(T_{\rm hi})} \right)^{1/3} \left( \frac{k}{k_0} \right)^{n_t} r.
\end{eqnarray}

Note that the GW spectrum is almost flat except for the weak
dependence on $k$ coming from $T_{\rm hi}(k)$ and the tensor spectral
index $n_t$ as long as the corresponding modes enter the horizon at
the RD era.

For completeness, we also present the GW spectrum for the long wavelength modes entering the horizon at the MD era:
\begin{equation}
	\Omega_{\rm GW}(t_0,k)
	= \frac{r}{12} \Omega_{\rm m,0}^2 \left( \frac{9H_0^2}{32k^2} \right) \Delta_{\rm {\cal R},prim}^2(k_0) \left( \frac{k}{k_0} \right)^{n_t},
\end{equation}
where $\Omega_{\rm m,0}$ is the density parameter of the matter at present.
By using $k_{\rm eq} = a_{\rm eq} H_{\rm eq} = \sqrt{2}\Omega_{\rm m,0}H_0/\sqrt{\Omega_{\rm r,0}} = 0.073\,\Omega_{\rm m,0} h^2$\,Mpc$^{-1}$, we thus obtain the interpolation formula from $k\ll k_{\rm eq}$ to $k\gg k_{\rm eq}$ as
\begin{equation}
	\Omega_{\rm GW}(t_0,k) \simeq \Omega_{\rm GW}(t_0,k \ll k_{\rm eq}) \left( \frac{g_{\ast}(T_{\rm hi})}{g_{\ast}(T_{\rm eq})} \right) \left( \frac{g_{\ast s}(T_{\rm eq})}{g_{\ast s}(T_{\rm hi})} \right)^{4/3} \left(1+ \frac{32 k^2}{9k_{\rm eq}^2} \right),
\end{equation}
which is equivalent to the expression given e.g. in Ref.~\cite{Nakayama:2009ce,Kuroyanagi:2011fy}.

\section{GW spectrum with a brief period of inflation}   
\label{sec:GW_inf}
\setcounter{equation}{0}

In this section we derive the GW spectrum with a brief period of
thermal inflation.

\subsection{GW spectrum with brief period of inflation}

First note that if the universe is dominated by some matter whose
equation of state is $w$, the total energy density and the Hubble
parameter evolves as
\begin{equation}
	\rho_{\rm tot}(t) \propto a(t)^{-3(1+w)} \to H \propto a(t)^{-3(1+w)/2}.
\end{equation}
Therefore, the scale factor at which the mode with wavenumber $k$ enters the horizon is given by
\begin{equation}
	k = a_{\rm in}(k)H_{\rm in}(k) \to a_{\rm in}(k) \propto k^{-2/(1+3w)}~~~{\rm for}~~~w>-1/3.
\end{equation}
For $w<-1/3$, the mode exits the horizon at
\begin{equation}
	k = a_{\rm out}(k)H_{\rm out}(k) \to a_{\rm out}(k) \propto k^{-2/(1+3w)}~~~{\rm for}~~~w<-1/3.
\end{equation}

\begin{figure}[tbp]
\begin{center}
\includegraphics[width=0.4\linewidth]{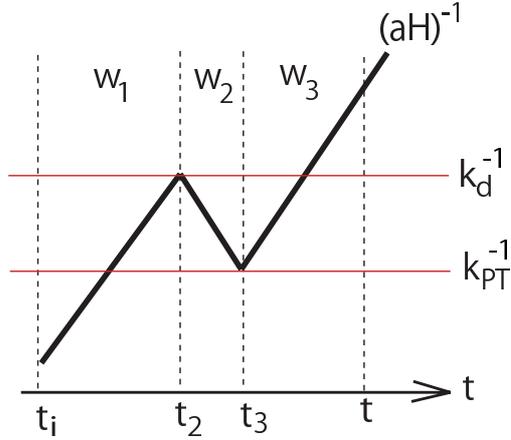}
\caption{
  Schematic picture for the evolution of the Hubble radius.
}
\label{fig:aH}
\end{center}
\end{figure}

Now let us consider the case where the equation of state changes as
$w_1 \to w_2 \to w_3$ with $w_1, w_3 > -1/3$ and $w_2 < -1/3$ as in
the case of thermal inflation (see Fig.~\ref{fig:aH}).
We want to evaluate the GW spectrum at late time $t\gg t_3$.

(i) $k<k_d$:
In the long-wavelength limit, we obtain
\begin{equation}
	\rho_{\rm GW}(k,t) \propto \rho_{\rm GW, prim}(k,t_i)\times \left( \frac{a(t_i)}{a_{\rm in}(k)} \right)^2
	\left( \frac{a_{\rm in}(k)}{a(t)} \right)^4
	\propto \rho_{\rm GW, prim}(k)\times k^{-4/(1+3w_3)}.
\end{equation}
Here 
\begin{eqnarray}
\rho_{\rm GW, prim}(k,t_i)
&\equiv& \frac{1}{32\pi G} \frac{k^2}{a(t_i)^2} \Delta_{\rm GW, prim}^2(k).
\end{eqnarray}

(ii) $k_d<k < k_{\rm PT}$:
In this case, the mode experiences the horizon entry and horizon exit, and again enters the horizon.
First, the spectrum at $t=t_2$ is given by
\begin{equation}
	\rho_{\rm GW}(k,t_2) \propto  \rho_{\rm GW, prim}(k,t_i)\times \left( \frac{a(t_i)}{a_{\rm in}(k)} \right)^2
	\left( \frac{a_{\rm in}(k)}{a(t_2)} \right)^4
	\propto \rho_{\rm GW, prim}(k)\times k^{-4/(1+3w_1)},
\end{equation}
Next, at $t=t_3$ it becomes
\begin{equation}
	\rho_{\rm GW}(k,t_3) \propto \rho_{\rm GW}(k,t_2) \times \left( \frac{a(t_2)}{a_{\rm out}(k)} \right)^4
	\left( \frac{a_{\rm out}(k)}{a(t_3)} \right)^2
	\propto \rho_{\rm GW}(k,t_2)\times k^{4/(1+3w_2)},
\end{equation}
Finally, at late time $t$, we have
\begin{equation}
	\rho_{\rm GW}(k,t) \propto \rho_{\rm GW}(k,t_3)\times \left( \frac{a(t_3)}{a_{\rm in}(k)} \right)^2
	\left( \frac{a_{\rm in}(k)}{a(t)} \right)^4
	\propto \rho_{\rm GW}(k,t_3)\times k^{-4/(1+3w_3)},
\end{equation}
In summary,
\begin{equation}
	\rho_{\rm GW}(k,t) \propto \rho_{\rm GW, prim}(k)\times 
	k^{-4/(1+3w_1)}  k^{4/(1+3w_2)} k^{-4/(1+3w_3)},
\end{equation}

(iii) $k>k_{\rm PT}$:
In the short-wavelength limit, we obtain
\begin{equation}
	\rho_{\rm GW}(k,t) \propto \rho_{\rm GW, prim}(k,t_i)\times \left( \frac{a(t_i)}{a_{\rm in}(k)} \right)^2
	\left( \frac{a_{\rm in}(k)}{a(t)} \right)^4
	\propto \rho_{\rm GW, prim}(k)\times k^{-4/(1+3w_1)}.
\end{equation}

\begin{figure}
  \centerline{\epsfxsize=9cm \epsfbox{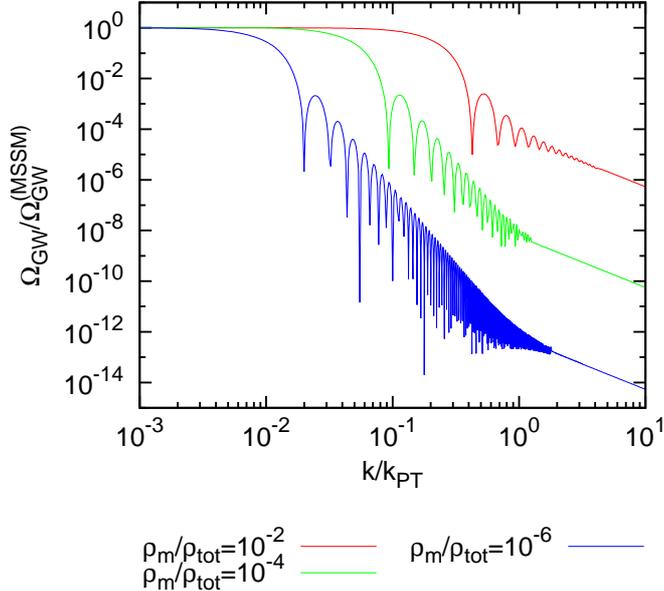}}
  \caption{\small GW spectrum with phase transition and instant decay
    into radiation.  We assumed that the universe is matter dominated
    before the vacuum energy dominates it, and varied the ratio of
    matter energy density to the total energy density at the phase
    transition.  }
  \label{fig_PT_MD}
\end{figure}

Collecting the above results, we finally arrive at
\begin{equation}
	\rho_{\rm GW}(k,t) \propto \left\{\begin{array}{ll}
	 \rho_{\rm GW, prim}(k)\times k^{-4/(1+3w_3)} &~~~{\rm for~~} k < k_d \\
	 \rho_{\rm GW, prim}(k)\times k^{-4/(1+3w_1)}  k^{4/(1+3w_2)} k^{-4/(1+3w_3)}
	 &~~~{\rm for~~} k_d < k < k_{\rm PT}\\
	 \rho_{\rm GW, prim}(k)\times k^{-4/(1+3w_1)} &~~~{\rm for~~} k > k_{\rm PT}
	\end{array}
	\right.
\end{equation}
Note that $\rho_{\rm GW, prim}(k) \propto k^2$. Here $k_d$ and $k_{\rm PT}$ are the
comoving Hubble scale at the beginning of thermal inflation and that
at the end of thermal inflation (or at the phase transition),
respectively (see Fig.~\ref{fig:aH}).  Thus in the case of
$w_1=w_3=1/3$ and $w_2=-1$, we obtain the GW spectrum
\begin{equation}
  \Omega_{\rm GW}(t_0, k) = \Omega_{\rm GW}^{\rm (std)}(t_0, k)
  \times \left\{\begin{array}{ll}
      1 &~~~{\rm for~~} k < k_d \\
      (k_d/k)^4 &~~~{\rm for~~} k_d < k < k_{\rm PT}\\
      (k_d/k_{\rm PT})^4 &~~~{\rm for~~} k > k_{\rm PT}
    \end{array}
  \right. .
\end{equation}
Therefore, in the intermediate frequency region, the spectrum is
proportional to $k^{-4}$ (see Fig.~\ref{fig_PT_RD_twin}).  Similarly,
for $w_1=0, w_3=1/3$ and $w_2=-1$, we have $\Omega_{\rm GW}(t_0, k)
\propto k^{-6}$ for $k_d < k < k_{\rm PT}$.  Fig.~\ref{fig_PT_MD}
shows the GW spectrum for the case where the universe is matter
dominated before thermal inflation and the vacuum energy instantly
goes to radiation after the phase transition, with varying the ratio
of matter energy density to the vacuum energy at the phase
transition. It is seen that the spectrum scales as $k^{-6}$ for $k_d <
k < k_{\rm PT}$.

Although the above arguments give global picture of the GW spectral shape,
actually there appear an oscillatory feature in the GW spectrum for $k_d < k < k_{\rm PT}$, 
which is not seen for $k < k_d$ and $k > k_{\rm PT}$ (see e.g., Fig.~\ref{fig_PT_RD_twin}).
Below we see more detail on the oscillatory feature of the GW spectrum.

\subsection{Oscillations in the GW spectrum}

As shown in \figref[fig_PT_RD_twin], the GW spectrum has typical oscillations in the case of brief period of thermal inflation.
Here we see the reason for such a behavior and derive the oscillation period.

\subsubsection{The reason for the oscillations in the GW spectrum}

The GW with modes $k_d < k < k_{\rm PT}$ experience additional
``horizon exit'' and ``horizon entry'' processes compared with other
modes (see Fig.~\ref{fig:aH}).  Let us denote the value of $h(t, k,
\lambda)$ and its time derivative at this horizon exit by $h(t_{\rm
  ex}, k, \lambda)$ and $\dot h(t_{\rm ex}, k, \lambda)$.  Note that,
although they take different values for different $k$ since they are
oscillating, the combination $[\dot h(t_{\rm ex}, k, \lambda)]^2 +
(k/a)^2 [h(t_{\rm ex}, k, \lambda)]^2$ is roughly the same for different
$k$.  After the horizon exit, the equation of motion of GWs is
approximated by
\begin{equation}
	\ddot h(t, k, \lambda) + 3H \dot h(t, k, \lambda) = 0,
\end{equation}
the general solution of which is given by
\begin{equation}
	h(t, k, \lambda) = H_1 + H_2 e^{-3Ht},
	\label{eq_GW_inf}
\end{equation}
where coefficients  $H_1$ and $H_2$ are determined by the boundary
condition at the horizon exit.  Then we find
\begin{eqnarray}
	h(t, k, \lambda) &=& \left( h(t_{\rm ex}, k, \lambda) + \frac{1}{3H} \dot h(t_{\rm ex}, k, \lambda)  \right)
	- \frac{1}{3H} \dot h(t_{\rm ex}, k, \lambda) e^{-3H(t-t_{\rm ex})} ,\\
	\dot h(t, k, \lambda)&=& \dot h(t_{\rm ex}, k, \lambda) e^{-3H(t-t_{\rm ex})}.
\end{eqnarray}
Therefore the part of the GW energy density proportional to $\dot h^2$
at the horizon exit soon damps.  The modes with $\dot h(t_{\rm ex}, k,
\lambda) = 0$ appear to be peaks in the GW spectrum.  On the other
hand, the modes satisfying $h(t_{\rm ex}, k, \lambda) +\dot h(t_{\rm
  ex}, k, \lambda)/3H = 0$ give troughs in the GW spectrum.  Thus we
expect oscillatory features in the GW spectrum for $k_d < k < k_{\rm
  PT}$ in the case of thermal inflation.

\subsubsection{Oscillation period}

Now we calculate the oscillation period in the GW spectrum.
We see that the oscillation period reflects the state of the universe before thermal inflation. 

First we consider the case where the universe was RD dominated before thermal inflation. We solve the Friedmann equation
\begin{eqnarray}
H^2 
= \frac{8 \pi G}{3} \left[ \rho_{\rm r,ref} \left( \frac{a_{\rm ref}}{a} \right)^4 + \rho_{\rm vac} \right],
\end{eqnarray}
where the subscript ``ref'' denotes some reference time before the phase transition, to get
\begin{eqnarray}
\frac{a}{a_{\rm ref}}
= \left( \frac{\rho_{\rm r,ref}}{\rho_{\rm vac}} \right)^{1/4} \sinh^{1/2} (2\omega_{\rm RD}t),
\end{eqnarray}
where 
\begin{eqnarray}
\omega_{\rm RD}
= \left( \frac{8 \pi G}{3} \rho_{\rm vac} \right)^{1/2}
\simeq H_{\rm PT}.
\end{eqnarray}
Here we have assumed that the inflation diluted the radiation enough. 
In the following we consider GWs which are well-inside around the vacuum-energy domination and 
well-outside the horizon at the phase transition. 
The variable $u \equiv k \int_0^t dt'/a(t')$ at the phase transition is
\begin{eqnarray}
u_{\rm PT}
&=& k \int_0^{t_{\rm PT}} \frac{dt}{a} \nonumber \\
&\simeq& \frac{1}{2} \frac{k}{a_{\rm PT} \omega_{\rm RD}} 
\left[ \frac{\rho_{\rm vac}}{(a_{\rm ref}/a_{\rm PT})^4 \rho_{\rm r,ref}} \right]^{1/4} \int_0^\infty \frac{dx}{(1+x^2)^{1/2}x^{1/2}} \nonumber \\
&\simeq& \frac{1}{2} \frac{k}{a_{\rm PT}H_{\rm PT}} f_{\rm r,PT}^{-1/4} \int_0^\infty \frac{dx}{(1+x^2)^{1/2}x^{1/2}},
\label{eq_uPT}
\end{eqnarray}
where $x=\sinh (2\omega_{\rm RD} t)$ and $f_{\rm r,PT} \equiv \rho_{\rm r,PT} / \rho_{\rm tot}$ is the energy fraction of radiation at the phase transition.
The oscillation peaks appear with the period of 
$\pi$ in $u_{\rm PT}$,\footnote{
As mentioned before, $h$ with wavenumbers of our interest goes out of the horizon during thermal inflation. 
One of the two solution to the equation of motion of GWs soon damp outside the horizon (see the previous subsection), 
and the peaks in the GW spectrum correspond to those wavenumbers which give maximal $H_1$ in \eqref[eq_GW_inf]. 
Such wavenumbers appear with period $\pi$ in terms of $u_{\rm he}$ (Here we denote ``horizon exit'' by ``he''). 
The point is that $u_{\rm he}$ is almost the same as $u_{\rm PT}$, 
since by the same calculation as \eqref[eq_uPT] we find
\begin{eqnarray*}
u_{\rm he}
= k \int_0^{t_{\rm he}} \frac{dt}{a}
= \frac{1}{2} \frac{k}{a_{\rm PT}H_{\rm PT}} f_{\rm r,PT}^{-1/4} \int_0^{x_{\rm he}} \frac{dx}{(1+x^2)^{1/2}x^{1/2}},
\end{eqnarray*}
with $x_{\rm he}$ given by
\begin{eqnarray*}
k=a_{\rm he} H_{\rm he} 
\leftrightarrow
\left( x_{\rm he} + \frac{1}{x_{\rm he}} \right)^{1/2}
\simeq \frac{k}{a_{\rm PT} H_{\rm PT}} \left( \frac{\rho_{\rm vac}}{\rho_{\rm r,PT}} \right)^{1/4}.
\end{eqnarray*}
One finds $x_{\rm he} \gg 1$ because both of the two factors $k/a_{\rm PT} H_{\rm PT}$ and 
$(\rho_{\rm vac} / \rho_{\rm r,PT})^{1/4}$ are much larger than 1. 
Note that $k/a_{\rm PT} H_{\rm PT} \gg 1$ although the GWs are out of horizon at the phase transition. 
}
therefore the gap between two neighboring peaks is
\begin{eqnarray}
\Delta u_{\rm PT}
= \frac{1}{2} \frac{\Delta k}{a_{\rm PT}H_{\rm PT}} f_{\rm r,PT}^{-1/4} \int_0^1 \frac{dx}{(1+x^2)^{1/2}x^{1/2}}
= \pi.
\end{eqnarray}
From this equation and $k_{\rm PT} \equiv a_{\rm PT} H_{\rm PT}$ we obtain
\begin{eqnarray}
\frac{\Delta k}{k_{\rm PT}} 
\simeq \frac{\pi}{1.854} f_{\rm r,PT}^{1/4}
\end{eqnarray}

In the case of MD universe before thermal inflation, we follow the same procedure. Solving the Friedmann equation
\begin{eqnarray}
H^2 
= \frac{8 \pi G}{3} \left[ \rho_{\rm m,ref} \left( \frac{a_{\rm ref}}{a} \right)^3 + \rho_{\rm vac} \right],
\end{eqnarray}
we get
\begin{eqnarray}
\frac{a}{a_{\rm ref}}
= \left( \frac{\rho_{\rm m,ref}}{\rho_{\rm vac}} \right)^{1/3} \sinh^{2/3} (2 \omega_{\rm MD} t),
\end{eqnarray}
where
\begin{eqnarray}
\omega_{\rm MD}
= \frac{3}{4} \left( \frac{8 \pi G}{3} \rho_{\rm vac} \right)^{1/2}
\simeq \frac{3}{4} H_{\rm PT}.
\end{eqnarray}
Then we get
\begin{eqnarray}
u_{\rm PT}
&=& k \int_0^{t_{\rm PT}} \frac{dt'}{a} \nonumber \\
&\simeq& \frac{1}{2} \frac{k}{a_{\rm PT} \omega_{\rm MD}} 
\left[ \frac{\rho_{\rm vac}}{(a_{\rm ref}/a_{\rm PT})^3 \rho_{\rm m,ref}} \right]^{1/3} \int_0^\infty \frac{dx}{(1+x^2)^{1/2}x^{2/3}} \nonumber \\
&\simeq& \frac{2}{3} \frac{k}{a_{\rm PT}H_{\rm PT}} f_{\rm m,PT}^{-1/3} \int_0^\infty \frac{dx}{(1+x^2)^{1/2}x^{2/3}},
\end{eqnarray}
and
\begin{eqnarray}
\frac{\Delta k}{k_{\rm PT}} 
\simeq \frac{\pi}{2.804} f_{\rm m,PT}^{1/3},
\end{eqnarray}
where $f_{\rm m,PT}$ is the energy fraction of matter at the phase transition.



\begin{thebibliography}{99}


\bibitem{Turner:1990rc} 
  M.~S.~Turner and F.~Wilczek,
  Phys.\ Rev.\ Lett.\  {\bf 65}, 3080 (1990).
  
\bibitem{Seto:2003kc} 
  N.~Seto and J.~'I.~Yokoyama,
  J.\ Phys.\ Soc.\ Jap.\  {\bf 72}, 3082 (2003)
  [gr-qc/0305096].
  
\bibitem{Tashiro:2003qp} 
  H.~Tashiro, T.~Chiba and M.~Sasaki,
  Class.\ Quant.\ Grav.\  {\bf 21}, 1761 (2004)
  [gr-qc/0307068].
    
\bibitem{Boyle:2005se} 
  L.~A.~Boyle and P.~J.~Steinhardt,
  Phys.\ Rev.\ D {\bf 77}, 063504 (2008)
  [astro-ph/0512014].
  
\bibitem{Boyle:2007zx} 
  L.~A.~Boyle and A.~Buonanno,
  Phys.\ Rev.\ D {\bf 78}, 043531 (2008)
  [arXiv:0708.2279 [astro-ph]].
    
\bibitem{Nakayama:2008ip} 
  K.~Nakayama, S.~Saito, Y.~Suwa and J.~'i.~Yokoyama,
  Phys.\ Rev.\ D {\bf 77}, 124001 (2008)
  [arXiv:0802.2452 [hep-ph]].
  
\bibitem{Nakayama:2008wy} 
  K.~Nakayama, S.~Saito, Y.~Suwa and J.~'i.~Yokoyama,
  JCAP {\bf 0806}, 020 (2008)
  [arXiv:0804.1827 [astro-ph]].
  
\bibitem{Kuroyanagi:2008ye} 
  S.~Kuroyanagi, T.~Chiba and N.~Sugiyama,
  Phys.\ Rev.\ D {\bf 79}, 103501 (2009)
  [arXiv:0804.3249 [astro-ph]].
    
\bibitem{Nakayama:2009ce} 
  K.~Nakayama and J.~'i.~Yokoyama,
  JCAP {\bf 1001}, 010 (2010)
  [arXiv:0910.0715 [astro-ph.CO]].
  
\bibitem{Nakayama:2010kt} 
  K.~Nakayama and F.~Takahashi,
  JCAP {\bf 1011}, 009 (2010)
  [arXiv:1008.2956 [hep-ph]].
  
\bibitem{Schettler:2010dp} 
  S.~Schettler, T.~Boeckel and J.~Schaffner-Bielich,
  Phys.\ Rev.\ D {\bf 83}, 064030 (2011)
  [arXiv:1010.4857 [astro-ph.CO]].
  
\bibitem{Durrer:2011bi} 
  R.~Durrer and J.~Hasenkamp,
  Phys.\ Rev.\ D {\bf 84}, 064027 (2011)
  [arXiv:1105.5283 [gr-qc]].
  
\bibitem{Kuroyanagi:2011fy} 
  S.~Kuroyanagi, K.~Nakayama and S.~Saito,
  Phys.\ Rev.\ D {\bf 84}, 123513 (2011)
  [arXiv:1110.4169 [astro-ph.CO]].
  
\bibitem{Saito:2012bb} 
  R.~Saito and S.~Shirai,
  Phys.\ Lett.\ B {\bf 713}, 237 (2012)
  [arXiv:1201.6589 [hep-ph]].

\bibitem{Seto:2001qf} 
  N.~Seto, S.~Kawamura and T.~Nakamura,
  Phys.\ Rev.\ Lett.\  {\bf 87}, 221103 (2001)
  [astro-ph/0108011];
  S.~Kawamura {\it et al.}, 
  Class.\ Quant.\ Grav.\  {\bf 28}, 094011 (2011).
  
\bibitem{Crowder:2005nr} 
  J.~Crowder and N.~J.~Cornish,
  Phys.\ Rev.\ D {\bf 72}, 083005 (2005)
  [gr-qc/0506015].

\bibitem{Cutler:2009qv} 
  C.~Cutler and D.~E.~Holz,
  Phys.\ Rev.\ D {\bf 80}, 104009 (2009)
  [arXiv:0906.3752 [astro-ph.CO]].
  
\bibitem{Yamamoto:1985rd} 
  K.~Yamamoto,
  Phys.\ Lett.\ B {\bf 168}, 341 (1986);
  G.~Lazarides, C.~Panagiotakopoulos and Q.~Shafi,
  Phys.\ Rev.\ Lett.\  {\bf 56}, 557 (1986).
  
\bibitem{Lyth:1995hj} 
  D.~H.~Lyth and E.~D.~Stewart,
  Phys.\ Rev.\ Lett.\  {\bf 75}, 201 (1995)
  [hep-ph/9502417];
  Phys.\ Rev.\ D {\bf 53}, 1784 (1996)
  [hep-ph/9510204].
  
  
\bibitem{Jinno:2011sw} 
  R.~Jinno, T.~Moroi and K.~Nakayama,
  Phys.\ Lett.\ B {\bf 713}, 129 (2012)
  [arXiv:1112.0084 [hep-ph]].
  
  
  
\bibitem{Weinberg:2003ur} 
  S.~Weinberg,
  Phys.\ Rev.\ D {\bf 69}, 023503 (2004)
  [astro-ph/0306304].
  
\bibitem{Dicus:2005rh} 
  D.~A.~Dicus and W.~W.~Repko,
  Phys.\ Rev.\ D {\bf 72}, 088302 (2005)
  [astro-ph/0509096].
  
\bibitem{Watanabe:2006qe} 
  Y.~Watanabe and E.~Komatsu,
  Phys.\ Rev.\ D {\bf 73}, 123515 (2006)
  [astro-ph/0604176].
  
\bibitem{Ichiki:2006rn} 
  K.~Ichiki, M.~Yamaguchi and J.~'I.~Yokoyama,
  Phys.\ Rev.\ D {\bf 75}, 084017 (2007)
  [hep-ph/0611121].
   
\bibitem{Jinno:2012xb} 
  R.~Jinno, T.~Moroi and K.~Nakayama,
  Phys.\ Rev.\ D {\bf 86}, 123502 (2012)
  [arXiv:1208.0184 [astro-ph.CO]].
    
  
  
  
\bibitem{Peccei:1977hh} 
  R.~D.~Peccei and H.~R.~Quinn,
  Phys.\ Rev.\ Lett.\  {\bf 38}, 1440 (1977).
  
\bibitem{Kim:1986ax} 
  For reviews, see J.~E.~Kim,
  Phys.\ Rept.\  {\bf 150}, 1 (1987);
  J.~E.~Kim and G.~Carosi,
  Rev.\ Mod.\ Phys.\  {\bf 82}, 557 (2010)
  [arXiv:0807.3125 [hep-ph]].
  
\bibitem{Kawasaki:2013ae} 
  For a review, see M.~Kawasaki and K.~Nakayama,
  arXiv:1301.1123 [hep-ph].

\bibitem{Hiramatsu:2010yn} 
  T.~Hiramatsu, M.~Kawasaki and K.~'i.~Saikawa,
  JCAP {\bf 1108}, 030 (2011)
  [arXiv:1012.4558 [astro-ph.CO]];
  T.~Hiramatsu, M.~Kawasaki, K.~'i.~Saikawa and T.~Sekiguchi,
  Phys.\ Rev.\ D {\bf 85}, 105020 (2012)
  [Erratum-ibid.\ D {\bf 86}, 089902 (2012)]
  [arXiv:1202.5851 [hep-ph]];
  JCAP {\bf 1301}, 001 (2013)
  [arXiv:1207.3166 [hep-ph]].

\bibitem{Kawasaki:2010gv} 
  M.~Kawasaki, N.~Kitajima and K.~Nakayama,
  Phys.\ Rev.\ D {\bf 82}, 123531 (2010)
  [arXiv:1008.5013 [hep-ph]];
  Phys.\ Rev.\ D {\bf 83}, 123521 (2011)
  [arXiv:1104.1262 [hep-ph]].
  
\bibitem{Moroi:2012vu} 
  T.~Moroi and M.~Takimoto,
  Phys.\ Lett.\ B {\bf 718}, 105 (2012)
  [arXiv:1207.4858 [hep-ph]].

\bibitem{Moroi:2013tea}
  T.~Moroi, K.~Mukaida, K.~Nakayama and M.~Takimoto,
  JHEP {\bf 1306} (2013) 040
  [arXiv:1304.6597 [hep-ph]].

\bibitem{Anisimov:2000wx} 
  A.~Anisimov and M.~Dine,
  Nucl.\ Phys.\ B {\bf 619}, 729 (2001)
  [hep-ph/0008058].

\bibitem{Laine:2010cq}
  M.~Laine,
  Prog.\ Theor.\ Phys.\ Suppl.\  {\bf 186}, 404  (2010)
  [arXiv:1007.2590 [hep-ph]].

\bibitem{Chikashige:1980ui} 
  Y.~Chikashige, R.~N.~Mohapatra and R.~D.~Peccei,
  Phys.\ Lett.\ B {\bf 98}, 265 (1981).
 
\bibitem{Chun:1994zp} 
  E.~J.~Chun, H.~B.~Kim and A.~Lukas,
  Phys.\ Lett.\ B {\bf 328}, 346 (1994)
  [hep-ph/9403217].

    \bibitem{seesaw}
T.~Yanagida, in Proceedings of the {\it{``Workshop on the Unified Theory and
 the Baryon Number in the Universe''}}, Tsukuba, Japan, Feb. 13-14, 1979, edited by
O.~Sawada and A.~Sugamoto, KEK report KEK-79-18, p. 95, 
and {\it{``Horizontal Symmetry And Masses Of Neutrinos''
}}, Prog. Theor. Phys. {\bf{64}} (1980) 1103;
M.~Gell-Mann, P.~Ramond and R.~Slansky, in {\it{``Supergravity''}}
 (North-Holland, Amsterdam, 1979) {\it{eds}}. D.~Z.~Freedom and P.~van 
Nieuwenhuizen, Print-80-0576 (CERN);
see also   P.~Minkowski,  Phys.\ Lett.\  B {\bf 67}, 421 (1977).

\bibitem{Maggiore:1999vm} 
  For a review, see M.~Maggiore,
  Phys.\ Rept.\  {\bf 331}, 283 (2000)
  [gr-qc/9909001].

\bibitem{Ade:2013zuv} 
  P.~A.~R.~Ade {\it et al.}  [Planck Collaboration],
  arXiv:1303.5076 [astro-ph.CO].

\bibitem{Croon:2013ana} 
  D.~Croon, J.~Ellis and N.~E.~Mavromatos,
  arXiv:1303.6253 [astro-ph.CO].
  
\bibitem{Nakayama:2013jka} 
  K.~Nakayama, F.~Takahashi and T.~T.~Yanagida,
  arXiv:1303.7315 [hep-ph];
  arXiv:1305.5099 [hep-ph].

\bibitem{Linde:1984cd} 
  A.~D.~Linde,
  Phys.\ Lett.\ B {\bf 132}, 317 (1983);
  C.~Destri, H.~J.~de Vega and N.~G.~Sanchez,
  Phys.\ Rev.\ D {\bf 77}, 043509 (2008)
  [astro-ph/0703417];
  R.~Kallosh and A.~D.~Linde,
  JCAP {\bf 0704}, 017 (2007)
  [arXiv:0704.0647 [hep-th]].

\bibitem{Bezrukov:2007ep} 
  F.~L.~Bezrukov and M.~Shaposhnikov,
  Phys.\ Lett.\ B {\bf 659}, 703 (2008)
  [arXiv:0710.3755 [hep-th]].
  
\bibitem{Ferrara:2010yw} 
  S.~Ferrara, R.~Kallosh, A.~Linde, A.~Marrani and A.~Van Proeyen,
  Phys.\ Rev.\ D {\bf 82}, 045003 (2010)
  [arXiv:1004.0712 [hep-th]];
  Phys.\ Rev.\ D {\bf 83}, 025008 (2011)
  [arXiv:1008.2942 [hep-th]].
  
\bibitem{Kallosh:2013pby} 
  R.~Kallosh and A.~Linde,
  arXiv:1306.3211 [hep-th].

\bibitem{Germani:2010gm} 
  C.~Germani and A.~Kehagias,
  Phys.\ Rev.\ Lett.\  {\bf 105}, 011302 (2010)
  [arXiv:1003.2635 [hep-ph]].
  
\bibitem{Nakayama:2010sk} 
  K.~Nakayama and F.~Takahashi,
  JCAP {\bf 1102}, 010 (2011)
  [arXiv:1008.4457 [hep-ph]].
  
\bibitem{Kamada:2012se} 
  K.~Kamada, T.~Kobayashi, T.~Takahashi, M.~Yamaguchi and J.~'i.~Yokoyama,
  Phys.\ Rev.\ D {\bf 86}, 023504 (2012)
  [arXiv:1203.4059 [hep-ph]].

\bibitem{Starobinsky:1980te} 
  A.~A.~Starobinsky,
  Phys.\ Lett.\ B {\bf 91}, 99 (1980).
  
\bibitem{Ellis:2013xoa} 
  J.~Ellis, D.~V.~Nanopoulos and K.~A.~Olive,
  arXiv:1305.1247 [hep-th].
  
\bibitem{Kallosh:2013lkr} 
  R.~Kallosh and A.~Linde,
  arXiv:1306.3214 [hep-th].

\bibitem{Linde:1994hy} 
  A.~D.~Linde,
  Phys.\ Lett.\ B {\bf 327}, 208 (1994)
  [astro-ph/9402031].
  
\bibitem{Vilenkin:1994pv} 
  A.~Vilenkin,
  Phys.\ Rev.\ Lett.\  {\bf 72}, 3137 (1994)
  [hep-th/9402085].
  
\bibitem{Harigaya:2012hn} 
  K.~Harigaya, M.~Kawasaki and T.~T.~Yanagida,
  Phys.\ Lett.\ B {\bf 719}, 126 (2013)
  [arXiv:1211.1770 [hep-ph]].
  
\bibitem{Buchmuller:2013lra} 
  W.~Buchmuller, V.~Domcke, K.~Kamada and K.~Schmitz,
  arXiv:1305.3392 [hep-ph].
  

\end{thebibliography}
\end{document}